\documentclass[12pt,preprint]{aastex}
\begin{document}
\def\ujy     {{$\mu Jy$}}
\def\Lya     {{Ly$\alpha$ }}
\def\Ha      {{H$\alpha$ }}

\title{The Stellar Masses and Star Formation Histories of Galaxies at
\boldmath $z\approx 6$: Constraints from {\it Spitzer} Observations in the Great
Observatories Origins Deep Survey
}

\author{Haojing Yan\altaffilmark{1},
Mark Dickinson\altaffilmark{2},
Mauro Giavalisco\altaffilmark{3},
Daniel Stern\altaffilmark{4},
Peter R. M. Eisenhardt\altaffilmark{4},
Henry C. Ferguson\altaffilmark{3}
}

\altaffiltext{1} {Spitzer Science Center, California Institute of Technology,
MS 220-6, Pasadena, CA 91125; yhj@ipac.caltech.edu}
\altaffiltext{2} {National Optical Astronomy Observatory, 950 N. Cherry St.,
Tucson, AZ 85719}
\altaffiltext{3} {Space Telescope Science Institute, 3700 San Martin Dr.,
Baltimore, MD 21218}
\altaffiltext{4} {Jet Propulsion Laboratory, California Institute of
Technology, 4800 Oak Grove Dr., Pasadena, CA 91109}

\begin{abstract}

   Using the deep {\it Spitzer} Infrared Array Camera (IRAC) observations of
the Great Observatories Origins Deep Survey (GOODS), we study the stellar
masses and star formation histories of galaxies at $z\approx 6$. The IRAC
instrument provides the best opportunity to estimate the stellar masses of
galaxies at these redshifts because it samples their rest-frame optical
fluxes, which are less prone to dust extinction and are more sensitive to the
light from longer-lived stars. Our study is based on the $i_{775}$-band dropout
sample selected from the GOODS southern and northern fields ($\sim$ 330
arcmin$^2$ in total), several of which already have spectroscopic confirmations.
In total, we derive stellar masses for 53 $i_{775}$-band dropouts that have
robust IRAC detections. These galaxies have typical stellar masses of 
$\sim 10^{10}M_\odot$ and typical ages of a couple of hundred million years,
consistent with earlier results based on a smaller sample of $z\approx 6$
galaxies in the {\it Hubble} Ultra Deep Field. 
We also study 79 $i_{775}$-band dropouts that are invisible in the IRAC data
and find that they are typically less massive by a factor of ten. 
These galaxies are much bluer than those detected by the
IRAC, indicating that their luminosities are dominated by stellar populations
with ages $\lesssim 40$ million years. We discuss various sources of
uncertainty in the mass estimates, and find that our results are rather robust.
The existence of galaxies as massive as $10^{10} M_\odot$ at $z\approx 6$ can
be explained by at least one set of N-body simulations of the hierarchical
paradigm. 
%However, the same simulations seem to result in a mass function that
%predicts more galaxies than we observe; even in the high-mass range where our
%IRAC data are rather complete, the model mass function predicts a factor of
%$\sim 6\times$ more objects. Possible sources of discrepancy are identified,
%but no definite answer can be reached at this stage. 
Based on our mass
estimates, we derive a lower limit to the global stellar mass density at 
$z\approx 6$. Considering the range of systematic uncertainties in the derived
stellar masses, this lower limit is 1.1 to 6.7$\times 10^6 M_\odot$~Mpc$^{-3}$
(co-moving), which is 0.2 to 1.1\% of the present-day value. The prospect of
detecting the progenitors of the most massive galaxies at yet higher redshifts
is explored: a deep,
wide-field near-IR survey using our current technology could possibly result in
positive detections at $z>7$. We also investigate the implication of our
results for reionzation, and find that the progenitors of the galaxies
comparable to those in our sample, even in the most optimized (probably
unrealistic) scenario, cannot sustain the reionization for a period longer than
$\sim 2$ million years. Thus most of the photons required for reionization must
have been provided by other sources, such as the progenitors of the dwarf
galaxies that are far below our current detection capability.

\end{abstract}


\section{Introduction}

   The Infrared Array Camera (IRAC; Fazio et al. 2004) of the 
{\it Spitzer Space Telescope} (Werner et al. 2004) has opened up a new window
for the study of galaxies at very high redshifts. Its impressive sensitivity
in the 3.6 and 4.5~$\mu$m channels enables the detection of galaxies as 
distant as $z\approx 6$, and thus, for the first time, offers the unique
opportunity of investigating the rest-frame optical properties of galaxies in
the early universe. At these wavelengths, the light from a galaxy is less 
affected by dust extinction and is more sensitive to the longer-lived stars
that dominate the stellar mass. For these reasons, IRAC observations can 
provide a wealth of information about the stellar population of galaxies at
very high redshifts, such as their ages and stellar masses. 

   The past year has witnessed substantial progress in this frontier.
Egami et al.\ (2005) used IRAC to detect a galaxy at $z \approx 6.7$
lensed by a foreground cluster and estimated that this object has a stellar
mass of $\sim 10^9 M_\odot$ and is at least $\sim$ 50~Myr old. Eyles et al.\
(2005) discussed two field galaxies at $z \approx 6$ that are detected in
the Great Observatories Origin Deep Survey (GOODS) IRAC data and found that
they are a few hundred Myr old and their stellar masses are a few 
$\times 10^{10} M_\odot$. Yan et al.\ (2005; Paper I) analyzed three galaxies
at $z\approx 6$ and 11 galaxies at $z\approx 5$ in the {\it Hubble} Ultra Deep
Field (HUDF) that are detected in the GOODS IRAC observations, and concluded
that the observed number density of massive galaxies
($\gtrsim 10^{10} M_\odot$) is consistent with predictions from contemporary
numerical simulations of galaxy formation in a cold dark matter ($\Lambda$CDM)
universe. Chary, Stern \& Eisenhardt (2005) investigated the red IRAC color of
a \Lya emitter at $z=6.56$, and suggested that the red color could be caused by
the presence of a very strong \Ha emission line (rest-frame equivalent width
$\sim 0.2\mu$m) redshifted to the $4.5\mu$m channel. Mobasher et al. (2005)
studied a galaxy in the HUDF that is not visible at optical wavelengths but is
very prominent in the IRAC images, and suggested that it could be an old galaxy
at $z\approx 6.5$ with a surprisingly high stellar mass of 
$5.7\times 10^{11} M_\odot$ (but see Chen \& Marzke (2004) and Yan et al.
(2004) for different interpretations of this same object).

   While some of the conclusions from these studies are still tentative, they
show that deep IRAC observations have enabled a new perspective on studies
of early galaxy formation. Previous studies, however, have been limited by the
small number of galaxies that were analyzed, and by the small fields of view,
where clustering variance might substantially affect space densities or other
properties of the galaxy samples. In this paper, we extend the analysis to the
combined areas of the two GOODS fields, in the regions of the {\it Hubble}
Deep Field North (HDF-N) and {\it Chandra} Deep Field South (CDF-S). These
fields cover roughly 30 times more solid angle than the HUDF alone, along two
sightlines. This substantially increases the number of galaxies we can study,
and should help average over variance due to large scale structure. We select
F775W-band dropouts (hereafter, $i_{775}$--dropouts) from the {\it HST}
Advanced Camera for Surveys (ACS) GOODS observations (Giavalisco et al. 2004a),
and identify counterparts from the full-depth, two-epoch {\it Spitzer} IRAC
observations of these fields (Dickinson et al., in preparation). These
candidate $z\approx 6$ galaxies are presented in \S 2 and their IRAC properties
are discussed in \S 3. We estimate their stellar masses in \S 4 and discuss the
implication of our results in \S 5. Our conclusions are summarized in \S 6.
The magnitudes quoted in this paper are all in the AB system. Throughout the
paper, we adopt the following cosmological parameters based on the first-year
{\it Wilkinson Microwave Anisotropy Probe} (WMAP) results from Spergel et al.\
(2003): $\Omega_M=0.27$, $\Omega_\Lambda=0.73$, and 
$H_0=71$\ km\ s$^{-1}$ Mpc$^{-1}$ (our results are not affected if using the
values from the three-year WMAP results for these parameters). All volumes
quoted are co-moving.

\section{Galaxy Candidates at \boldmath $z\approx$ 6 in the GOODS Fields}

   The $z\approx 6$ galaxy candidates used in this paper are the 
$i_{775}$-dropouts selected from the complete, 5-epoch GOODS ACS observations
in both the HDF-N and CDF-S (Giavalisco et al. 2004a). The selection of these
candidates will be discussed in detail by Giavalisco et al. (2006, in 
preparation). Briefly, the selection is aimed at the redshift range of 
$5.5\lesssim z\lesssim 6.5$, and requires a candidate have a red color across
the F750W ($i_{775}$) and F850LP ($z_{850}$) bands, and be invisible in the
F606W ($V_{606}$) and F435W ($B_{435}$) bands. 

    Quantitatively, the 
candidates must meet these criteria: $S/N\geq 5$ in $z_{850}$, 
$(i_{775}-z_{850})>1.3$ mag, and $S/N<2$ in both the $V_{606}$ and $B_{435}$.
These criteria are the same as have been previously used by the GOODS team
(Giavalisco et al. 2004b; Dickinson et al. 2004). While the colors
are calculated based on the SExtractor (Bertin \& Arount 1996) MAG\_ISO
magnitudes, MAG\_AUTO magnitudes are used when quoting the $z_{850}$ magnitudes
of these candidates as they are closer to the ``total'' magnitudes. 
As brown dwarfs in our Galaxy have similar red colors, an ``image sharpness''
criterion, based on comparing peak $z_{850}$-band surface brightness within a
$0\farcs125$ aperture to the isophotal magnitude, was used to reject possible
point sources from the $i_{775}$-dropout sample down to $z_{850} = 26.5$ mag.
This criterion was purposefully tuned to be conservative: it may improperly
reject a few galaxies, but should successfully eliminate most stars down to
that magnitude limit. At fainter magnitudes, it becomes difficult to robustly
distinguish stars and compact galaxies in the GOODS ACS images, as the 
stellarity loci of stars and galaxies
%, which are well separated at brighter fluxes, 
start to merge for any of the criteria we have considered. 

   After visual inspection to eliminate likely spurious sources, the resulting
sample consists of 274 $i_{775}$--dropout candidates (145 in the south, 129 in
the north). Of these, eight more were further rejected as possible stars on the
basis of having full-width-at-half-maximum (FWHM)~$< 0\farcs125$; two of these
have been spectroscopically confirmed as stars by Vanzella et al.\ (2006).
After this process, 142 and 124 valid candidates remain in the CDF-S and HDF-N, 
respectively. 

   We have identified five close pairs among these $i_{775}$-dropouts, three
in the CDF-S sample and two in the HDF-N sample, respectively. The separation
of the components in these pairs are all smaller than 2$''$, which means that
these pairs, if detected in the IRAC images, will be detected only
as single sources (see below). Therefore, for the statistical purpose of this
paper, we take the number of $i_{775}$-dropouts in the CDF-S and HDF-N as 139
and 122, respectively.

\section{IRAC Detections of \boldmath $z\approx 6$ Galaxy Candidates}

   The IRAC data used in this paper are the mosaics of the full, two-epoch
data of the GOODS observations in both the CDF-S and HDF-N (Dickinson et al.,
in preparation; see also
http://data.spitzer.caltech.edu/popular/goods/Documents/goods\_dataproducts.html).
The mosaics cover approximately $10^{'}\times 16^{'}$
in each field (total coverage of $\sim 330$ arcmin$^2$), and have a nominal
exposure time of $\sim$ 23.2 hours in each of the 3.6, 4.5, 5.8, and 8.0
$\mu$m channels. Because of the way the observations were designed, the
exposure time in the central $\sim$ 20\% stripe of each field is twice as long.
The final pixel size of the mosaics is 0.6$^{''}$~pixel$^{-1}$, i.e., roughly
half of
the native IRAC pixel size. Sources were detected in a weighted sum of the 3.6
and 4.5~$\mu$m images using SExtractor, and their magnitudes were measured in
each band through circular apertures that are 3.0$^{''}$ in diameter. These
aperture magnitudes were converted to total magnitudes after adding aperture
corrections (see Paper I). We only include sources with $S/N>3$ (measured in
3.6$\mu$m within the above mentioned aperture) for cross-matching.

   The $i_{775}$-dropouts were cross-matched with the IRAC sources
following the procedures adopted in Yan et al. (2004; 2005). As the FWHM of the
point spread function (PSF) in the 3.6$\mu$m channel is about 1.8$''$, the
matching was done with a
2$''$ search radius. This radius was chosen partly because of the close pairs
in our sample, whose IRAC centroids are offset from those of the individual
components. The matched sources were then visually inspected to ensure that the
identifications were secure. In order to avoid ambiguity in interpreting the
measured fluxes, we excluded any IRAC sources that are blended with 
neighbouring objects. 

   In total, 64 $i_{775}$-dropouts have thus been securely identified in the
IRAC data, with 40 in the CDF-S and the remaining 24 in the HDF-N. Among these
sources, there are two pairs in the CDF-S and one pair in the HDF-N, 
respectively, all of which are counted as single IRAC objects in the numbers 
quoted above. Except for these three pairs, all other objects have their IRAC
and ACS centroids matched to within $0.6^{''}$, which is fully consistent with
our previous experience (see Yan et al. 2004; 2005).

   The Lyman-break selection of galaxies at $z\approx 6$ is known to suffer
some contamination from low-redshift, early-type galaxies that have similar red
colors. Our IRAC data can be used to further reduce these contaminators. A
number of the $i_{775}$-dropouts identified above have very bright counterparts
in the IRAC images, giving flux density ratios 
$f_\nu(3.6\mu m)/f_\nu(z_{850})>20$ (or equivalently, 
$(z_{850} - m_{3.6\mu m})>3.25$ mag). Thus they satisfy the
selection criterion of the ``IRAC-selected Extremely Red Objects'' (IEROs; Yan
et al. 2004), which are likely old, passively-evolving galaxies at 
$z\approx 2.4$. Therefore, these objects are excluded from our final sample.
There are six and five such IEROs among the IRAC-detected $i_{775}$-dropouts in
the CDF-S and HDF-N, respectively. The ratio of contamination due to IEROs is
15\% for the CDF-S (six out of 40) and 21\% for the HDF-N (five out of 24). 

   We also consider the $i_{775}$-dropouts that are not detected by the
GOODS IRAC observations. We visually examined the IRAC images at the 
positions of the $i_{775}$-dropouts that were not cross-matched in the above
procedure and found that 45 objects in the CDF-S and 34 objects in the HDF-N
were undetected because of their faintness --- quantitatively, they all have
$S/N<3$ in both the 3.6 and 4.5~$\mu$m images. The remaining 118 objects were
not cross-matched because they are blended with nearby, unrelated sources.

   To summarize, we are left with a final sample of 250 objects for analysis
(133 in the CDF-S and 117 in the HDF-N). We have counted the close pairs as
single sources, and have excluded the possible contaminators (brown dwarfs and
IEROs). The final sample of $i_{775}$-dropouts that are
securely detected in the IRAC data has a total of 53 objects (34 in the CDF-S
and 19 in the HDF-N). The 3.6~$\mu$m-band magnitudes of these objects range
from 23.27 to 26.50 mag. The final sample of $i_{775}$-dropouts undetected in
the IRAC data consists of 79 objects (45 in the CDF-S and 34 in the HDF-N).
In the rest of the paper, we will refer to these two samples as the 
``IRAC-detected sample'' and the ``IRAC-invisible sample''. The remaining 118
$i_{775}$-dropouts (54 in the CDF-S and 64 in the HDF-N) are blended with
unrelated neighbours. IEROs make up $\sim$ 17\% of the non-blended sample, and
it is plausible that the same fraction also holds in the blended sample.
Therefore, a total of 98 objects in the blended sample could be galaxies at
$z\approx 6$ after excluding this fraction of IERO contamination. We will refer
to the sample of these 98 objects as the ``blended sample''.

   Despite that we have utilized every means to eliminate various sources of
contamination, our final sample could still have some contaminators. This is
particularly possible among the objects of low $S/N$, where spurious sources
are more likely to occur and low-redshift, red galaxies (not necessarily as
extreme as the IEROs) are more easily scattered above the color selection
threshold. To address this issue, we use the much deeper ACS data in the HUDF
to assess the quality of our candidate selection. Thirteen objects in our CDF-S
candidate list are within the coverage of the HUDF, and all of them are
verified to be real objects. Therefore, the contamination caused by spurious
sources seems to be negligible to the $S/N$ level that we consider. However,
three of these thirteen sources are not among the HUDF $i_{775}$-dropout list
of Yan \& Windhorst (2004b). These three sources ($S/N\approx 5$ in the GOODS
data) have $i_{775}-z_{850} < 0.5$ mag in the deeper HUDF data, indicating that
the contamination rate
in our final sample caused by photometric error could be as high as 
$\sim$ 23\%. We will take this into account when discussing the implications of
our results in \S 5.

   Fig. 1 shows the $z_{850}$-band magnitude distributions of the IRAC-detected
and IRAC-invisible samples as histograms of different colors. Their median 
magnitudes are 26.32 and 27.00 mag, respectively. For comparison, the median
magnitude of the blended sample is 26.71 mag. While the objects in the
IRAC-invisible sample are fainter on average, the majority of them are still
within the range of the IRAC-detected sample.

\section{Constraining the Stellar Masses of \boldmath $z\approx 6$ Galaxies}

   In this section, we constrain the stellar masses of $z\approx 6$ galaxies in
both the IRAC-detected and the IRAC-invisible samples as defined
in \S 3. For this purpose, we compare our observations to the stellar
population synthesis models of Bruzual \& Charlot (2003; hereafter BC03).
As per Paper I, we explore the models of the following star formation 
histories (SFHs): instantaneous bursts (or Simple Stellar Populations, SSPs),
and continuous bursts with exponentially declining star formation rate (SFR) in
the form of $SFR\propto \tau^{-1}$~$exp(t/\tau)$, where the time-scale $\tau$
ranges from $\tau = 10$~Myr to 1~Gyr. The step size in $\tau$ is 10~Myr for 
$\tau =10$~Myr to 0.1~Gyr, and is 0.1~Gyr for $\tau=0.1$ to 1~Gyr. For each of
these SFHs, we generate models with ages ranging from 1~Myr to 1~Gyr; the step
size in age is 10~Myr when the age is less than 0.1~Gyr, and is 0.1~Gyr when it
ranges from 0.1 to 1~Gyr. To facilitate comparison to other, published studies,
we use a Salpeter initial mass function (IMF; Salpeter 1955) extending from 0.1
to 100 $M_\odot$. Many recent studies indicate that the actual IMF in most
environments probably has fewer low-mass stars ($< 1 M_\odot$) than are
predicted by the Salpeter IMF power-law slope (e.g., Kroupa 2001; Chabrier 
2003). These stars contribute very little to the observed light from galaxies,
and therefore, to first order, changing the low-mass IMF simply rescales all
masses and star formation rates derived from comparison to stellar population
models. Generally speaking, using a Chabrier IMF would lower the mass estimates
by about 50--60\%. Changes to the IMF shape and slope at higher masses,
however, may lead to wavelength-dependent and time-dependent changes in the
mass-to-light ratios ($M/L$) relative to the assumed Salpeter models.

   In Paper I, we carried out detailed SED analysis using models of different
metallicities and various reddening values, and allowed the redshift be a
free parameter when it was unavailable. When single-component models did not
fit well, we also considered two-component models. Here, however, it is 
difficult to take this approach, because the objects in the IRAC-detected 
sample are only significantly detected in two to three passbands ($z_{850}$ and
3.6$\mu$m, some are also detected in 4.5$\mu$m). This is particularly true for
the IRAC-invisible objects, as they are only significantly detected in one band
($z_{850}$). Therefore, we have to use a different method, detailed below. We
will mainly concentrate on models of solar metallicity and zero reddening, but
will also discuss the effects of different metallicities and reddening values. 

  Currently, seven objects in the IRAC-detected and -invisible samples have
already been spectroscopically confirmed at $z>5.5$ (Stanway et al. 2004a,
2004b; Dickinson et al. 2004; Vanzella et al. 2006; Stern et al., in 
preparation; Vanzella et al., in preparation). For the sake
of simplicity, however, we choose to assign a common redshift of $z=6$ for all
objects. We will discuss the bias caused by this simplification in 
\S 4.3 together with other sources of uncertainty.

\subsection {Stellar Masses of IRAC-Detected Objects}

   For each object, we first examine how its luminosity in the IRAC bands
constrains the upper limit of its stellar mass. The luminosity of an object
with fixed mass is determined by its age and SFH, and becomes fainter as it
gets older; i.e., its $M/L$ becomes larger as it evolves. It can reach the
largest possible $M/L$ if it forms all its stars through an instantaneous burst
and passively evolves to its maximal age (i.e., a SSP; see Fig. 2). Therefore,
an object can have the highest possible mass (assuming no dust extinction) if
its luminosity is dominated by such a ``maximally old", single-burst component.
This methodology is similar to that adopted in Papovich et al. (2001).
In practice, we assume that all the flux of an object detected in the 
3.6$\mu$m channel comes from this maximally old component. The 3.6$\mu$m 
channel is chosen because it has the highest sensitivity of the IRAC bands, and
thus the measurements in this channel have the highest S/N. 
As we assume the objects in our sample are all at $z=6$, their maximal age
allowed (i.e., setting the formation redshift $z_f=\infty$)
in our adopted cosmology is 0.95~Gyr. To match the model grid in age domain,
we take their maximal ages as 1.0~Gyr. The stellar mass upper limit of
a given object is then obtained by comparing its 3.6$\mu$m magnitude to the 
prediction given by the 1.0~Gyr-old SSP model.
Hereafter we refer to this limit as ``$M_{\rm max}$'', and the method as
the ``maximally-old component method"
\footnote{Assuming no dust extinction (discussed in \S 4.3), the $M_{\rm max}$
value thus obtained marks a firm upper limit to the mass. The maximally-old
component always underpredicts the flux in $z_{850}$-band, and to explain the
observed $z_{850}$-band flux would require a secondary, young component be
added. In such a two-component, ``old+young'' model, the derived mass of a
galaxy will be smaller than the $M_{\rm max}$ value that we obtained, because
the contribution from the young component will reduce the $M/L$. The exact
amount of such a reduction depends on the assumed age and SFH of the young
component; using the suite of models shown in Fig. 3 for this young component,
the derived maximum mass will be smaller than our current $M_{\rm max}$ value
by a few to 50\% on average. As we have no further constraints on the young
component, we choose to adopt the $M_{\rm max}$ value based on our current
simplistic model as a safe upper limit.
}. 

   Next, we find the lower bound to the stellar mass (hereafter
$M_{\rm min}$) of each object. If the SFH of an object is known, its 
($z_{850}-m_{3.6\mu m}$) color can be used as an indicator of its age (see
Fig. 3). Its mass can be derived through comparison of the observed 3.6$\mu$m
magnitude to the prediction of the corresponding model at the inferred age. As
we do not know what SFH a given object has, we consider the full range of
models and obtain one mass estimate for each model. For each galaxy, the 
minimum among all these estimates is taken as its $M_{\rm min}$.

   Given our assumptions, the true mass of a galaxy will lie 
between $M_{\rm min}$ and $M_{\rm max}$. While there is no other information
that can further help us judge what its true mass might be, we use the median
mass from the set of exponential SFH models fitted to match the 
$(z_{850}-m_{3.6\mu m})$ color. We refer to this as the ``representative mass''
of this galaxy (hereafter $M_{\rm rep}$). We also refer to the age corresponding
to this $M_{\rm rep}$ as its ``representative age'' ($T_{\rm rep}$). The 
representative masses range from 0.09--7.0$\times 10^{10} M_\odot$ (the median
is 9.5$\times 10^9 M_\odot$) and the representative ages range from 50--400 Myr
(the median is 290 Myr); both are consistent with one of the general
conclusions in Paper I that some galaxies as massive as a few 
$\times 10^{10} M_\odot$ were already in place at $z\approx 6$, and that they
are typically a few hundred million years old. The histograms of the three 
values, $M_{\rm min}$, $M_{\rm rep}$ and
$M_{\rm max}$, are shown in the top panel of Fig. 4. As a further justification
of using $M_{\rm rep}$ and $T_{\rm rep}$, we note that $M_{\rm rep}$
for J033240.01-274815.0 is 2.1$\times$10$^{10}M_\odot$, which is well within a
factor of two of the best-fit mass of 3.4$\times$10$^{10}M_\odot$ that Paper I
derived for this object using a more sophisticated SED analysis. 
For comparison, the $T_{\rm rep}$ value of this object is 200 Myr, while the 
analysis of Paper I shows that the evolved component of this object has an age
of 500 Myr.

\subsection{Stellar Masses of IRAC-Invisible Objects}

   The fact that many objects are undetected in the IRAC data immediately
suggests that they are likely much less massive that the IRAC-detected ones.
The 3.6$\mu$m flux density upper limits of these sources, measured as
2~$\sigma$ fluctuation within a 3$^{''}$-diameter aperture, range from 0.029 to
0.091~$\mu$Jy (27.73 to 26.50 mag), with the median of 0.042~$\mu$Jy
(27.35 mag). Using the maximally-old-component method described above, we
derive the stellar mass upper limits of these objects based on their flux
density upper limits. Again, we assume all galaxies are at $z=6$. These upper
limits, which are shown in the dotted histogram displayed in the lower panel of
Fig. 4, range from 3.0--9.4$\times 10^{9} M_\odot$. The median of these
{\it upper limits} is 4.3$\times 10^{9} M_\odot$, which is not only less than
the upper limit of the IRAC-detected sample in general, but is also 
significantly less than the median of the {\it representative mass} of the
latter by a factor of two.

   Fig. 5 compares the ($z_{850} - m_{3.6\mu m}$) colors of the IRAC-detected
objects (red squares) with the upper limits of these colors of the 
IRAC-invisible objects (downward arrows)\footnote{The ``gap'' in this figure
that seemingly seperates the IRAC-detected and IRAC-invisible objects does not
imply that these are two distinctly different populations. Instead, the
transition between these two samples should be rather smooth. The ``gap'' is
largely caused by the sources that fall just slightly below our detection
threshold in IRAC ($S/N>3$), whose ($z_{850} - m_{3.6\mu m}$) color limits
are then calculated by using their 2~$\sigma$ flux upper limits in the
$3.6\mu m$-band.}. This comparison suggests that the
IRAC-invisible objects are not only less massive than the IRAC-detected ones
but are also bluer on average, indicating that young stellar populations
are playing a dominant role --- by luminosity --- in these galaxies.

   To further test the robustness of the conclusion that these objects are
less massive, we stacked their 3.6$\mu$m images to increase the S/N of the
detection. For each object, a $12.6^{''}\times12.6^{''}$ section (i.e., 
21-pixel on a side) around its center was cut out from the $3.6\mu$m image.
We then combined these image ``stamps'' together by taking their median value
at each pixel. We adopted median rather than average as the stacking algorithm,
because the former is more effective in rejecting the contaminating pixels from
the neighbours around invisible sources. The final median stack is 
displayed in the left panel of Fig. 6, which shows a visible, albeit weak,
source at the center (see pixels within the circle). For comparison, the
right panel of this figure shows the stack of the same number of randomly
chosen image stamps, where no detectable source at the center can be seen. 

   This weak source represents the average property of the IRAC-invisible 
$i_{775}$-dropouts in the 3.6$\mu m$ channel. To extract its flux, we utilized
a $1.2{''}$-diameter (2 pixels) aperture. The background value and the aperture
correction to total flux were determined through simulation. The simulation was
performed for 20 runs, and in each run 830 different artificial galaxies with
a total magnitude of 27.50 mag were randomly distributed over the real image.
Image stamps of these artificial galaxies (16,600 in total) were cut out and 
stacked in the same way as we stacked the IRAC-invisible sources. The peak value
of the pixel histogram of this stack was chosen as the background value, and the
magnitude of the stacked artificial galaxy was extracted using a 
$1.2{''}$-diameter aperture. The aperture correction was calculated as the
difference between this aperture magnitude and the input total magnitude.
The same background value and the aperture correction were then applied to the
real stack of the IRAC-invisible object. Its total magnitude
thus obtained is 27.44 mag, which is consistent with the flux density upper
limits described above.

   By the same token, the representative property of this median IRAC source 
in $z_{850}$-band can be described by the median $z_{850}$ magnitude of the
IRAC-invisible sample, which is 27.00 mag. Using the same analysis in \S 4.1,
we obtain $(M_{\rm min}, M_{\rm rep}, M_{\rm max})$ estimates for a typical
source in the IRAC-invisible sample as
$(1.5\times 10^8, 2.0\times 10^8, 5.9\times 10^9) M_\odot$. For all three types
of estimates, the masses derived from the median-stack photometry are much
smaller than those for the objects in the IRAC-detected sample (see the
dashed lines in the top panel of Fig. 4). The $T_{\rm rep}$ value of this
median source is 30~Myr.
 
\subsection{Sources of Uncertainty in Mass Estimates}

  A major source of uncertainty is in the systematic of the measurement of the
($z_{850}-m_{3.6\mu m}$) color. As it is used as the age estimator, this color
affects both $M_{\rm min}$ and $M_{\rm rep}$ (but not $M_{\rm max}$). In this
study, we use total magnitude for $m_{3.6\mu m}$ and SExtractor MAG\_AUTO for
$z_{850}$. Although MAG\_AUTO is frequently taken to represent the total 
magnitude of a galaxy, in practice it may be subject to biases, particularly
for faint objects in high-resolution {\it HST} images. Monte Carlo simulations
of photometry for artificial objects in the GOODS ACS images indicates that at
the typical magnitudes of our $i_{775}$--dropouts, the MAG\_AUTO magnitude
underestimates the total
$z_{850}$-band flux of an object by $\sim$ 0.5 mag. This means that the true 
($z_{850}-m_{3.6\mu m}$) color could be 0.5 mag bluer than our current
estimate, and thus the age of a given object could be younger, and 
hence the $M/L$ could be smaller. We find that our current $M_{\rm min}$ and 
$M_{\rm rep}$ values could have been overestimated by $\sim$ 21\% if the
the ($z_{850}-m_{3.6\mu m}$) color is indeed 0.5 mag bluer than we adopted.

  In addition to photometry, there are more complicated sources of uncertainty.
When deriving the results shown in previous sections, we have made three major 
simplifications, namely, the galaxies (or equivalently, the models) are all at
$z=6$, all have solar abundance and all are free of dust obscuration. Here
we discuss the systematics introduced by such simplifications. We mainly 
discuss the effects on the derived stellar mass, as our current study largely
concentrates on this quantity. 
We consider how the derived masses would change if we vary the assumed values
for the redshift, metallicity, and reddening.
We express the systematics in terms of relative difference in mass, 
$\Delta M^i/M$, where the superscript $i$ can be ``$z$'' (redshift), 
``Fe/H'' (metallicity) and ``red'' (reddening). The quantity $\Delta M$ is
defined as $\Delta M = M-M^{'}$, where $M$ is the value obtained using our
adopted models, and $M^{'}$ is the value obtained when the model parameter
in question is changed. A positive value of $\Delta M^i/M$ indicates that our
default models yield a larger estimate for the mass than those where parameter
$i$ is changed, and a negative value means the opposite. We consider the
effects of model parameter changes on all three mass estimates considered here,
namely, $M_{\rm min}$, $M_{\rm rep}$, and $M_{\rm max}$. We only discuss the
IRAC-detected sample, but the conclusions could similarly be applied to the
IRAC-invisible sample.

    By fixing the redshift, we ignore the differences in the redshifted SED,
in the luminosity distance, and in the amount of the IGM
absorption. The first two factors (especially the luminosity distance) are 
the most relevant in deriving $M_{\rm max}$ using a maximally-old SSP. Using a
$z=6$ SSP will overestimate the $M_{\rm max}$ values for objects at lower 
redshifts, and will underestimate them for objects at higher redshifts.
This is demonstrated in the bottom panel of Fig. 7 for two extreme cases of
$z=5.5$ (asterisks) and $z=6.5$ (open squares). The $M_{\rm max}$ values of our
objects in the IRAC-detected sample have been recalculated using a SSP at
$z=5.5$ and a SSP at $z=6.5$, respectively. Our original values
would overestimate by 21\% if the objects were actually all at $z=5.5$,
and would underestimate by 27\% if they were actually all at $z=6.5$.

  For galaxies at $z \gtrsim 5.9$, the \Lya forest suppresses the flux detected
in the $z_{850}$-band. This has no effect on the maximal mass estimates, 
$M_{\rm max}$, which is based only on the flux in the 3.6$\mu$m channel.
However, it does impact the mass estimates $M_{\rm min}$ and $M_{\rm rep}$,
which make use of $M/L$ derived from the $z_{850}$-3.6$\mu$m color. For example,
if the object under question is actually at $z>6$, its observed 
($z_{850} - m_{3.6\mu m}$) color would be redder than what is expected when
at $z=6$, because its flux in the $z_{850}$-band is more severely suppressed.
Applying a $z=6$ template to such a color will result in an artificially older
age (see Fig. 3), and hence an artificially larger $M/L$ (see Fig. 2). As a
result, both $M_{\rm min}$ and $M_{\rm rep}$ will be overestimated. This
overestimate, however, is partially canceled by the change in the luminosity 
distance and the $k$-correction. The top and middle panels of Fig. 7 show that
the assumption of $z=6$ would overestimate $M_{\rm min}$ and $M_{\rm rep}$ by
about 25\% for galaxies that were in fact at $z = 6.5$ (see the open squares).
 At $z < 6$, this effect results in underestimates. However, as the change of
the IGM opacity across the $z_{850}$-band is negligible at this redshift
regime, the mass estimates change by only a few percent
if the galaxies were instead at $z = 5.5$. This is shown in the top and middle
panels of Fig. 7 as asterisks.

   In Paper I, we considered the metallicities varying from solar abundance to
1/200 of solar, the latter being the smallest value available from the BC03 
models. We found that the $z\approx 6$ objects in the HUDF are usually best
fitted by models with solar abundance. For this reason, we have adopted the
simplifying assumption of solar metallicity for our fiducial models.
As it is not likely that objects in the early universe would have higher
abundance, here we only examine the extreme case at the low metallicity end,
i.e., how our results would be biased if all the objects actually have their
abundance of 1/200 of solar. Fig. 8 shows the relative differences of the three
mass estimates, all evaluated at $z=6$. Models with subsolar metallicites
generally have bluer colors at a fixed age; hence, older ages are required to
match the observed colors, and this leads to larger $M/L$. Therefore, the 
masses based on the colors ($M_{\rm min}$ and $M_{\rm rep}$) derived using
the solar abundance models are smaller than those derived from models with 
$Z = Z_\odot /200$ by 70\% and 32\%, on average. For $M_{\rm max}$, instead,
the solar abundance models have higher $M/L$, and hence the masses are 32\% 
larger than those for models with 1/200th solar metallicity.

    Dust reddening and obscuration will also affect the derived masses. In Paper
I, we found that the best-fit models always have near-zero reddening, although
an amount up to $E(B-V) = 0.4$ to 0.5 mag could not be excluded. This is
because we had to account for the very blue rest-frame UV colors observed for
the galaxies studied in that paper, which we did by invoking two-component
models with both younger and older stars. The best-fit models always have
$E(B-V) \approx 0$ if we require both components to have the same reddening.
However, if this {\it ad hoc} requirement is removed, the reddening becomes
unconstrained. Here, we consider the effects of non-zero reddening on our
results. For mass estimates based on a color measurement, i.e., $M_{\rm min}$
and $M_{\rm rep}$, the effects of reddening and age on $M/L$ derived from the
models tend to cancel, resulting in relatively small changes in the final
derived masses. Dust makes the observed 3.6$\mu$m flux fainter, thus increasing
$M/L$. However, it also reddens the spectrum, requiring a younger model to
match the observed ($z_{850} - 3.6\mu$m) color, and this younger model will
have an intrinsically smaller $M/L$ (before applying the extinction).
For illustration, we consider the effects of $E(B-V) = 0.2$ mag of
extinction, assuming the attenuation law of Calzetti et al.\ (2000). The
results are shown in Fig. 9.
Although this reddening corresponds to nearly 1 mag of extinction in the
observed 3.6$\mu$m band and more than 2 mag at $z_{850}$, the average change in
the derived $M_{\rm min}$ and $M_{\rm rep}$ is only 13\% and 30\%
(overestimated),
respectively, as the model age changes to compensate the extinction.
$M_{\rm max}$, on the other hand, is derived from the photometry at 3.6$\mu$m
alone, and hence changes by a factor of 3.3 (underestimated). 

   In the end, we find that most of our mass estimates are relatively robust
against large changes in the values assumed for the redshift, metallicity, and
extinction.  In most cases, varying those parameters over the ranges described
here, the derived masses change by less than 35\%. If metallicity were as low
as 1/200th solar, then our values for $M_{\rm min}$ (derived assuming 
$Z_\odot$ models) would be underestimated by 70\% on average. The largest
effect would be that of reddening on our estimates of $M_{\rm max}$, which
could be increased by a large factor if reddening were substantial. However, we
note that this would imply substantial reddening on the old (1 Gyr) stellar
population assumed in the toy model from which the mass upper limits are
derived.  In practice, the mass could be increased arbitrarily by adding dust
to an old stellar population, but such a model seems rather arbitrary and
unphysical, and also fails to take into account the observed UV light and
UV-to-optical rest-frame colors of the galaxies. Instead, the mass estimates
derived from the observed colors ($M_{\rm min}$ and $M_{\rm rep}$) are 
comparatively insensitive to reddening. Other stellar population properties are
more sensitive to these systematic effects. In particular, both age and SFR
are completely degenerate with extinction when only a single color is used for
analysis. Thus, while we have shown that the mass estimates are comparatively
robust, other properties are more uncertain, and rely more heavily on the
inference for low extinction from more detailed analysis of the smaller HUDF
sample in Paper I.

\subsection{Legitimacy of Adopted Declining SFHs}

   The stellar mass estimates that we obtained in this study hinge upon the
exponentially declining SFHs that we assumed. Such SFHs imply that the 
star-formation activities of these objects were more intense in the past than
in the epoch when they are observed. While it is impossible to directly verify
this picture based on our current data, we can at least test if our results are
self-consistent with the assumption that the SFHs of these objects are a
declining function of time.

   One quantity that can be used to address this question is ``specific SFR'',
which is defined as the ratio of the current SFR and the total stellar mass.
The reciprocal of this quantity is the period of time (denoted as 
$\widetilde{T}$) during which a galaxy could build up its total mass if it were
maintaining the same, constant SFR in the past. The current SFR can be 
calculated according to eqn. 2 of Madau, Pozzetti \& Dickinson (1998),
i.e., $L_{UV}=8.0\times 10^{27}\times SFR$, where $L_{UV}$ is the flux (in
ergs/s/Hz) at rest-frame 1500\AA. The SFR thus derived is insensitive to the
past SFH when the age of the galaxy is much longer than the life time of O and
B stars at the main sequence. We convert the $z_{850}$-band 
magnitudes to $L_{UV}$, again assuming that the objects are all at $z=6$, and
ignoring the caveat that the $z_{850}$-band actually samples the continuum at
around 1300\AA. 

    Fig. 10 shows the specific SFR (calculated by dividing the $M_{\rm rep}$
into the SFR value) of the IRAC-detected sample as the function of
stellar mass (solid triangles). The result for the median-stack of the 
IRAC-invisible objects is also plotted (the asterisk). The data points
form a tight sequence that is between constant SFR of 3.0 and 
30~$M_\odot yr^{-1}$,
which are shown as the two straight lines to the left and right, respectively.
This primarily reflects the rest-frame UV magnitude range
of the sample ($24.9 < z_{850} < 27.4$ mag) and its conversion into
star formation rates ($30 > {\mathrm SFR}/(M_\odot {\mathrm yr^{-1}}) > 3$).
The lower envelope in specific star formation rates corresponds to the
$z_{850}$ magnitude limit of the sample.  The upper envelope may
have greater physical significance, corresponding to the 
high-luminosity cut-off in the UV luminosity function for galaxies at $z
\approx 6$.

   The $\widetilde{T}$ values are much larger than the representative galaxy
ages, $T_{\rm rep}$, obtained in the previous section. More than half of these 
objects (58\%) have $\widetilde{T}$ larger than the age of the universe at
$z\approx 6$ (some are even $>4$ Gyr), which is a clear indication that
there would not have been sufficient time for them to acquire their masses if
their past SFRs were as low as their current values. In other words, their
SFRs must have been declining and must be much larger in the past, which is 
consistent with the assumed models.

   One caveat of the discussion above, however, is the possible role of dust
extinction. If the galaxies in our sample suffered a reddening of $E(B-V)=0.2$
mag (as discussed in \S 4.3), their true ongoing SFR would be higher than
the current estimates by a factor of $\sim 8.5$, and yet their $M_{\rm rep}$
values would be lower by $\sim$ 30\% (see \S 4.4). As a result, their
$\widetilde{T}$ values would be smaller by a factor of $\sim 12$, and hence
the majority of them could still build up their stellar masses within the
time allowed by the age of the universe. However, even in this case, the
$\widetilde{T}$ values (with the median of 155~Myr) are still much larger than
$T_{\rm rep}$ derived based on the reddened models (with the median of 10~Myr). 
Therefore, this is still consistent with our assumption that the objects
in our sample have higher SFRs in the past.

\section{Cosmological Implications}

   The results presented above show that some galaxies as massive as a few
$\times 10^{10} M_\odot$ were already in place by $z\approx 6$, and that their
typical ages are a few hundred million years old, which are consistent with
our earlier results in Paper I. In this section, we discuss the
cosmological implications of these results.

\subsection{Comparison to $\Lambda$CDM Simulations}

%% In Paper I, SPH: 2.1e-4, overestimated by a factor of 6x
%%             TVD: 3.7e-4, overestimated by a factor of 3.8x
%% Model: > 1.6e10 M_sun
%%      SPH -- 3.464e-5
%%      TVD -- either 2 to 3 objects; meaning 9.66e-5; 
%% Observed: >1.6e10 M_sun
%%  M_rep:
%%    before blending-correction:  2.6875e-5
%%    after  blending-correction:  1.74*2.6875e-5=4.676e-5 (observed/sph: 1.35)
%%  M_min: 
%%    before                       1.500e-5
%%    after                        1.74*1.500e-5=2.61e-5 (observed/sph: 0.75)
%%  M_max:
%%    before                       6.125e-5
%%    after                        1.74*6.125e-5=10.658e-5 (obs./sph: 3.077)
%%%

%%%  TVD: 9.66e-5/2.61e-5=3.7; 9.66e-5/4.676e-5=2.1; 9.66e-5/10.658e-5=0.9

%   Paper I has shown that the number density of very massive galaxies at
%$z\approx 6$ is consistent with the mass function (MF) predicted by at least
%one set of numerical simulations of the hierarchical paradigm (Nagamine et al.
%2004; Night et al. 2006). 
%but using the much larger sample obtained in our current

   Following Paper I, here we further investigate how well the observed number
density of very massive galaxies at $z\approx 6$ can be explained by the
numerical simulations of the hierarchical paradigm. As in Paper I, we make
comparison to the mass functions (MF) predicted by two types of hydrodynamic
simulations in a $\Lambda$CDM universe, namely, a Smoothed Particle 
Hydrodynamics (SPH) simulation and a Total Variation Diminishing (TVD)
simulations (Nagamine et al. 2004; Night et al. 2006). The simulation box sizes
are 100$h^{-1}$ Mpc and 22$h^{-1}$ Mpc, respectively. The cumulative MF from
these two simulations are given in Fig. 11. As the simulation volume is not
unlimited, both MF are cut-off at the high-mass end (at around 
$1.8\times 10^{11}$ and $2.0\times 10^{10} M_\odot$ for the SPH and TVD models,
respectively). Nevertheless, comparing to these models is still meaningful, as
they extend to the regime where our observations are rather complete. For
reasons as described in detail in Paper I, our IRAC data should be complete for
detecting galaxies with $M>1.6\times 10^{10} M_\odot$, which is indicated in
the figure as the vertical dotted line.

    In Paper I, we derived a lower limit for the cumulative number
density above the same mass threshold, which is 8.0$\times 10^{-5}$~Mpc$^{-3}$,
and we concluded that the predictions from the SPH and TVD simulations were a
factor of 2.6$\times$ and 4.6$\times$ lager than this lower limit. However, we
recently discovered that the mass functions from the SPH and TVD simulations
were presented with incorrect normalization in Paper I,
leading to predicted number densities that were
too high by a factor of 6.0 and 3.8, respectively. Using the corrected numbers,
the SPH and TVD predictions should be 0.4$\times$ and 1.2$\times$ of the
observed lower limit. 

    As the sample in Paper I is very small (three objects), the much larger
sample in our current study offers a better statistics for comparison.
Using the redshift selection function for our color criteria, which
we evaluate from Monte Carlo simulations following the procedures described in
Giavalisco et al. (2004b), we obtain the effective volume of 8.0$\times 10^5$
Mpc$^3$ for our $i_{775}$-dropout sample. The cumulative number densities
inferred from the
($M_{\rm min}, M_{\rm rep}, M_{\rm max}$) values of the IRAC-detected sample
are shown as the blue, yellow and red solid curves, respectively. 
If we assume that the stellar mass distribution of the blended sample is the
same as the non-blended sample, we can obtain the total contribution from the
combined non-blended and blended samples by scaling the results that we
obtained for the non-blended sample. This scaling factor is 1.74. The 
dot-dashed curves in Fig. 11 show the observed number density after this
correction being applied, and are what will be used to compare to the models.
The flattening of the observed curves at lower masses is a consequence of
increasing incompleteness at fainter magnitudes.

    At the mass threshold of $M>1.6\times 10^{10} M_\odot$, the cumulative
number density derived from our observations is
(2.6, 4.7, 10.6)~$\times 10^{-5}$Mpc$^{-3}$ based on
($M_{\rm min}, M_{\rm rep}, M_{\rm max}$), respectively. The SPH model
prediction is $\sim 1.3\times$, 0.7$\times$ and 0.3$\times$ of the
observations. The TVD model prediction, on the other hand, is $\sim 3.7\times$,
2.1$\times$ and 0.9$\times$ of the observations. This comparison shows that
the number density of the observed high mass galaxies agrees resonably well
with the models to within the uncertainty of the stellar mass estimates. The
prediction from the TVD model is considerably higher than that from the SPH
model, which reflects the uncertainty in simulations caused by the differences
in the implementation methods, such as the mass and spatial resolutions, the
simulation box sizes, and the initial conditions (Nagamine, priv. 
communications). In particular, the smaller box size of the TVD simulation has
a significant impact at the high-mass end, as the statistics is dominated by
only a small number of objects (only 2--3 galaxies above our choosen mass
threshold). Nevertheless, the comparison between our observations and these
simulations does show that the models of the hierachical paradigm
are capable of producing sufficient number of high mass galaxies at the
early epoch stage of the universe.

\subsection{Global Stellar Mass Density}

   Based on the stellar masses obtained for our sample, here we estimate
the global stellar mass density at $z\approx 6$. To address this
question, ideally one would need a mass-limited, complete sample. However, the
$i_{775}$-dropout sample that we have been using is not mass-limited. Since
it was selected from the rest-frame UV luminosities of the objects,
which do not correlate directly with stellar masses. While
the incompleteness of this sample can be quantified as a function of rest-frame
UV luminosity, this cannot easily be transferred to the mass-domain. In fact,
the discussion in \S 5.1 suggests that our sample could possibly suffer from
significant incompleteness even at the high-mass end. Nevertheless,
we can still use our current sample to constrain the lower limit of the global
stellar mass density at $z\approx 6$. 

   Adding the ($M_{\rm min}, M_{\rm rep}, M_{\rm max}$) estimates of the 
individual objects in the IRAC-detected samples, the total stellar mass density
locked in these galaxies 
is (5.8, 9.1, 31.4)$\times 10^5 M_\odot$Mpc$^{-3}$. The 
contribution inferred from the IRAC-invisible samples is much smaller. The
total stellar mass contributed by this sample can be approximated by 
multiplying the total number of objects to the stellar mass estimates of the
median stack described in \S 4.2, 
which corresponds to stellar mass density of 
(0.14, 0.20, 5.9)$\times 10^5 M_\odot$Mpc$^{-3}$. Therefore, the
sum of both samples is (5.9, 9.3, 37.3)$\times 10^5 M_\odot$Mpc$^{-3}$.
We also correct for the incompleteness caused by crowding by considering the
IRAC-blended sample in the same way as in \S 5.1 (i.e., multiplying by a factor
of 1.74). The final stellar mass density after this correction is
(10.4,16.3,64.8)$\times 10^5 M_\odot$Mpc$^{-3}$.

  This result is shown in Fig. 12, together with the estimates at lower
redshifts (Dickinson et al. 2003). The data are shown in terms of ratios to the
critical mass density, which is 1.4$\times 10^{11} M_\odot$Mpc$^{-3}$ in our
adopted cosmology. We note that, however, these results have not yet taken into
account the fact that our $i_{775}$-dropout sample could still be contaminated
by low-redshift galaxies that are not as extreme as the IEROs (see \S 3).
Considering this effect, the results presented above could be reduced 
by $\sim$ 23\%.

   Finally, we emphasize that, while the objects in our IRAC-invisible
sample make a small contribution to the total stellar mass
density, this does not necessarily mean that the less massive galaxy
population only makes up an insignificant fraction of the total stellar mass
density at $z\approx 6$. If the MF of galaxies at $z\approx 6$ has a
very steep slope at the low-mass end (see Fig. 11), the contribution from the
less massive galaxies that are missing from our IRAC data could be much higher
than that from the galaxies at the high-mass end.

\subsection {Progenitors of High-mass Galaxies: Detectability at \boldmath $z>7$}

   In \S 4.4, we argue that the objects in the IRAC-detected sample should have
much higher SFRs in the past. This implies that the progenitors of these
objects were once much more UV-luminous than they are at $z\approx 6$. Since
there have already been some efforts to push the current redshift boundary to
$z>7$ (e.g., Bouwens et al. 2004), it is relevant to discuss the prospect of
detecting such early galaxies from a different perspective.

   Consider a typical object in the high-mass subset
%   As an example, here we discuss a typical object in the high-mass subset
%($M>10^{10}M_\odot$) 
of the IRAC-detected sample that has 
$M_{\rm rep}=2.0\times 10^{10} M_\odot$. The stellar mass and the instantaneous
SFR that its progenitor has at a given age depend on its SFH, as does its
luminosity. Using the BC03 models that we adopt, we predict the apparent
magnitude of its progenitor at any given redshift for a range of formation and
observed redshifts. Fig. 13 shows two examples by assuming $z_f=7.8$ 
(top panel) and $z_f=9.0$ (bottom panel), and gives its apparent magnitude in
$J$-band if it is observed 50~Myr (blue) and 100~Myr (red) after the birth. We
choose these two formation redshifts for demonstration because the age of the
object when evolving to $z\approx 6$ would be $\sim$ 200--400~Myr, which is
within the range of $T_{\rm rep}$ of the objects in the high-mass subset. In 
most models that we consider, the progenitor of such a very massive object
should be detected at $J\approx 24$ mag. Based on the number of similar objects
in the IRAC-detected sample, the surface density of such progenitors
would be $\sim 0.05$ arcmin$^{-2}$. While it is too low for the existing deep,
pencil-beam surveys using the Near Infrared Camera and Multi Object 
Spectrometer onboard the {\it HST} (e.g., Thompson et al. 1999, 2005; Dickinson
et al. 2000; Bouwens et al. 2004, 2005), such a surface density is high enough
to produce a significant number of detections through a deep survey over a few
hundred arcmin$^2$ that is now feasible using the new generation of wide-field
near-IR instruments at ground-based observatories.

   This simple argument echos the ones made about three decades ago regarding
the progenitors of the local giant ellipticals, when it was suggested that such 
progenitors (``primeval galaxies'') could have formed most of their stars
within a short period of time and thus their rest-frame UV luminosities could
be so high that they could be easily detected at very high redshifts (e.g.,
Partridge \& Peebles 1967; Sunyaev, Tinsley \& Meier 1978; Shull \& Silk 1979).
Even at that time, however, it was realized that dust extinction could easily
quench the UV fluxes and thus those progenitors might not be as bright as 
expected (e.g., Kaufman 1976). The same is also true in our discussion here.
For example, a modest amount of dust equivalent to a reddening of 
$E(B-V)\approx 0.2$ mag could easily suppress our predicted $J$-band 
brightness by more than 2 magnitudes, and therefore would make the detection of
such progenitors at $z>7$ much more difficult. A null detection from a deep,
wide-field survey would still be valuable, however, as it could then be used to
constrain how dusty the very early galaxies could be and shed new light to
their forming mechanism.

   There is another factor that could affect the detectability of the
progenitors of high-mass galaxies. If the high-mass objects that we see at 
$z\approx 6$ are the merger products of many much less massive progenitors, the
luminosity of an individual progenitor could be much fainter and escape
detection. In this case, a null detection would imply a high merger rate within
the first several hundred million years of the universe, which would provide
some constraints on the hierarchical formation theory.

\subsection{Implication for Reionization}

   As it is possible that the high-mass galaxies were born through very
intense star-formation processes, it is interesting to study the role of
their progenitors in reionizing the neutral hydrogen in the early universe.
Specifically, here we investigate if their progenitors could have provided 
sufficient ionizing photons to sustain the reionization. To maximize their
contribution to the ionizing photon budget, we consider an extreme -- probably
unphysical -- case where {\it all} such high-mass galaxies were formed at the
same time at $z_{\rm reion}$, which is the redshift when the reionization began.

   The precise value of $z_{\rm reion}$ is still not known. Assuming
instantaneous reionization with ionization fraction of $x_e=1$ at 
$z<z_{\rm reion}$, Kogut et al. (2003) found $z_{\rm reion}=17\pm 3$ based on
the first-year WMAP results. Allowing for uncertainties in the reionization
history, Kogut et al. obtained $11<z_{reion}<30$ (at 95\% confidence level).
The most recent, three-year WMAP results have refined this estimate, but the
derived value of $z_{\rm reion}$ is still model-dependent
(Spergel et al. 2006). Assuming that the reionization happened instantaneously
at $z>7$, they found the maximum-likelihood value of $z_{\rm reion}=11.3$. For
the purpose of demonstration, we use $z_{\rm reion}=9.0$, 11.3 and 15.0 as
examples in the discussion below.

   According to Madau, Haardt \& Rees (1999; their Eqn. 27), the critical SFR
density required to reionize the universe can be calculated as
$\rho_{SFR}^{cri}(z) \approx 0.013 f_{esc}^{-1} \times [(1+z)/6]^3$~$M_\odot$~yr$^{-1}$~Mpc$^{-3}$ (assuming a Salpeter IMF).
This quantity, calculated by assuming $f_{esc}=0.1$, is shown as the black,
solid curves in Fig. 14 for the three assumed $z_{\rm reion}$ values. The 
contribution from the progenitors of the high-mass
galaxies to the total SFR density depends on their SFH, and is less if the SFH
is more prolonged. Fig. 14 shows the results for three toy models of 
exponentially declining SFHs with $\tau$ of 0.1 (blue), 1.0 (green) and 10 Myr
(red). The solid, dashed and dotted curves represent the results based on the
minimum, representative and maximum global stellar mass densities shown in Fig.
12, respectively. These calculations demonstrate that, unless the bulk of their
stars are {\it simultaneously} formed through an intense burst of very short
duration ($\tau\lesssim 10$~Myr), the contribution from the progenitors of the
galaxies similar to those in our sample is not sufficient to reionize the 
universe. Even if they indeed came into being through an unphysically short,
simultaneous burst with $\tau\lesssim 10$~Myr, they cannot keep the universe
ionized for long (only $\lesssim 3$~Myr). 

   This is not to say, however, that star-forming galaxies cannot reionize the
universe, because here we only count the progenitors of galaxies that have
comparable stellar masses to those in our sample. If the MF has a very steep
slope at the low-mass end, dwarf galaxies (which are below the detectability of
our current observations) can contribute sufficient amount of
ionizing photons, similar to the arguments made by Yan \& Windhorst (2004a, b)
for the case at $z\approx 6$.

\section{Summary}

   In this paper, we use the full-epoch IRAC observations of the GOODS 
{\it Spitzer} Legacy Program to study the stellar masses of galaxies at 
$z\approx 6$. In total, 53 $i_{775}$-dropouts (seven of them with spectroscopic
confirmation) selected in the entire GOODS fields ($\sim 330$ arcmin$^2$) are
securely detected by the IRAC. We derive the stellar masses for all these
objects, using three different mass estimators: $M_{\rm min}$, $M_{\rm rep}$,
and $M_{\rm max}$. We argue that the true mass of an object should be between
$M_{\rm min}$ and $M_{\rm max}$, and use $M_{\rm rep}$ as its ``representative
mass''. We further refer to the age to which $M_{\rm rep}$ corresponds as its
``representative age'', $T_{\rm rep}$. We find that the $M_{\rm rep}$ values
range from 0.09--7.0$\times 10^{10} M_\odot$, and the $T_{\rm rep}$ values
range from 50--400 Myr, both are consistent with the results of our earlier
work based on a smaller sample in the HUDF (Paper I). We also study 79
$i_{775}$-dropouts that are invisible from the IRAC observations, and find that
they are generally one order of magnitude less massive than those detected by
IRAC. Their ACS-to-IRAC colors are much bluer than those of the IRAC-detected
ones, indicating that their luminosities are dominated by young population of
unreddened stars with ages $\lesssim 40$~Myr. We discuss various sources of
uncertainty in the mass estimates, and find that most of our mass estimates are
relatively robust.

   We discuss a number of implications of our results. As in Paper I, we find
that the existence of the most massive galaxies in our much enlarged sample can
still be explained by at least one set of N-body simulations of the 
hierarchical paradigm. 
%The much enlarged sample in this study enables us to compare the
%observed number densities against the mass function predicted by these
%simulations. We find that the models generally predict a factor of six times
%higher number density even in the mass range where our IRAC data should be
%rather complete. 
We derive a lower limit to the global stellar mass density
at $z\approx 6$, and conclude that at least 0.2--1.1\% of the total stellar
mass in the local universe has been locked in stars by $z\approx 6$. As the
most massive galaxies in our sample should have much larger SFRs in the past,
we discuss the prospect of detecting their progenitors at $z>7$. A near-IR
survey covering a few hundred arcmin$^2$ to a depth of $AB\sim 24.0$ mag
can provide valuable constraints on the properties of such progenitors even
in the case of a null detection. We also discuss how the progenitors of galaxies
comparable to those in our sample could contribute in reionizing the hydrogen 
in the early universe, and conclude that such progenitors alone are not
sufficient to sustain the reionization of the universe.

\acknowledgments

We thank the other members of the GOODS team who have contributed to the 
success of the observations and data analysis. We also thank our referee,
Rodger Thompson, for his prompt and insightful referee report. We are grateful
to Kentaro Nagamine for his up-to-date MF. We wish to thank Ivo Labbe for
useful discussion.
Support for this work, part of the {\it Spitzer Space Telescope} Legacy Science
Program, was provided by NASA through Contract Number 1224666 issued by the Jet
Propulsion Laboratory, California Institute of Technology under NASA contract
1407.  HY acknowledges the support from the NASA grant HST-GO-09780.03.

%\newpage

\begin{figure}
%\plotone{fig_idrop_hist.eps}
\plotone{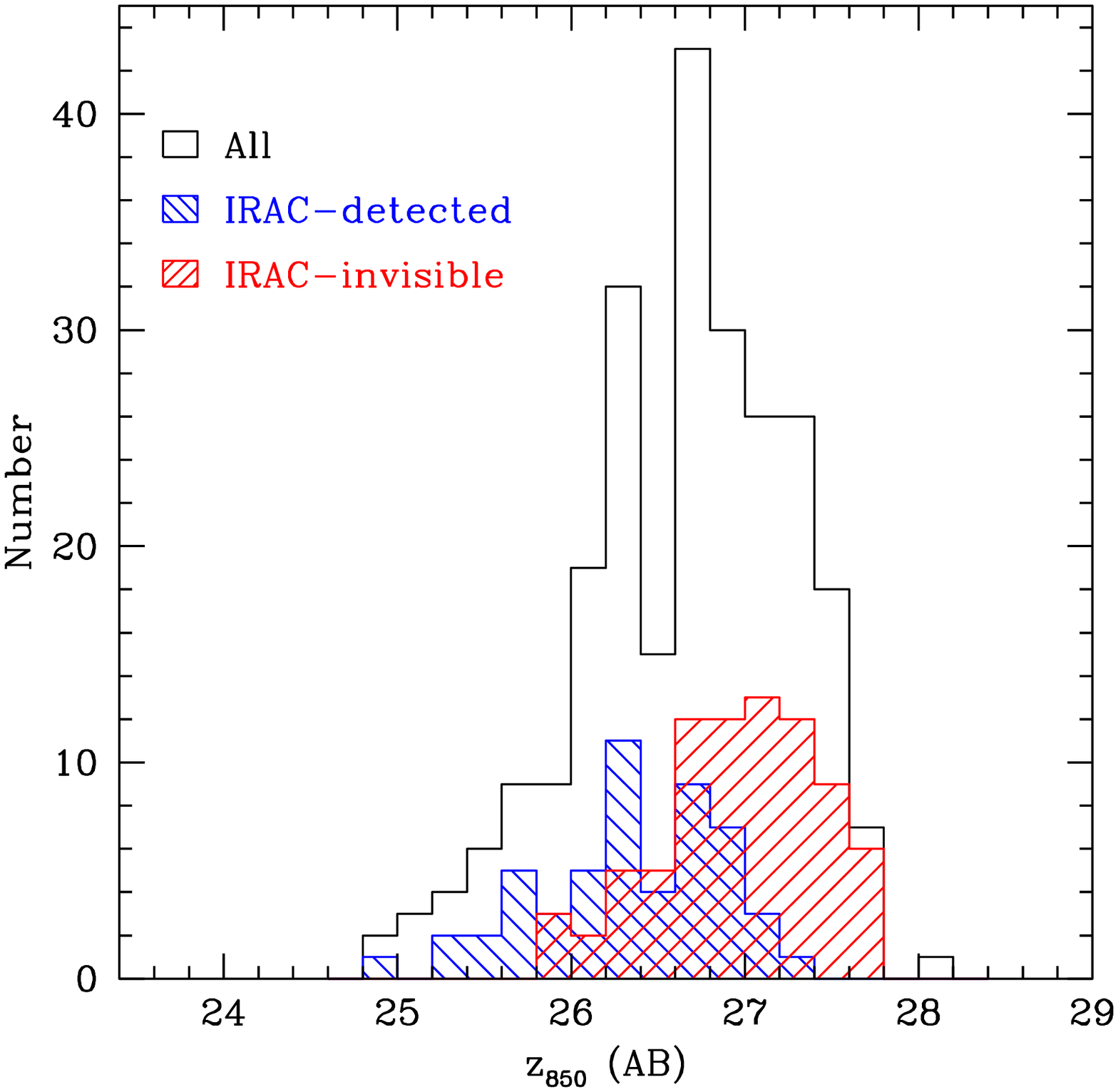}
\caption{The distribution of $z_{850}$ magnitudes of the $i_{775}$-dropouts in
the GOODS fields. The shaded blue and red histograms are for the IRAC-detected
and IRAC-invisible samples, respectively, while the black, unfilled histograms
are for all objects (including the blended sample).
}
\end{figure}

\begin{figure}
%\plotone{fig_age_ML.eps}
\plotone{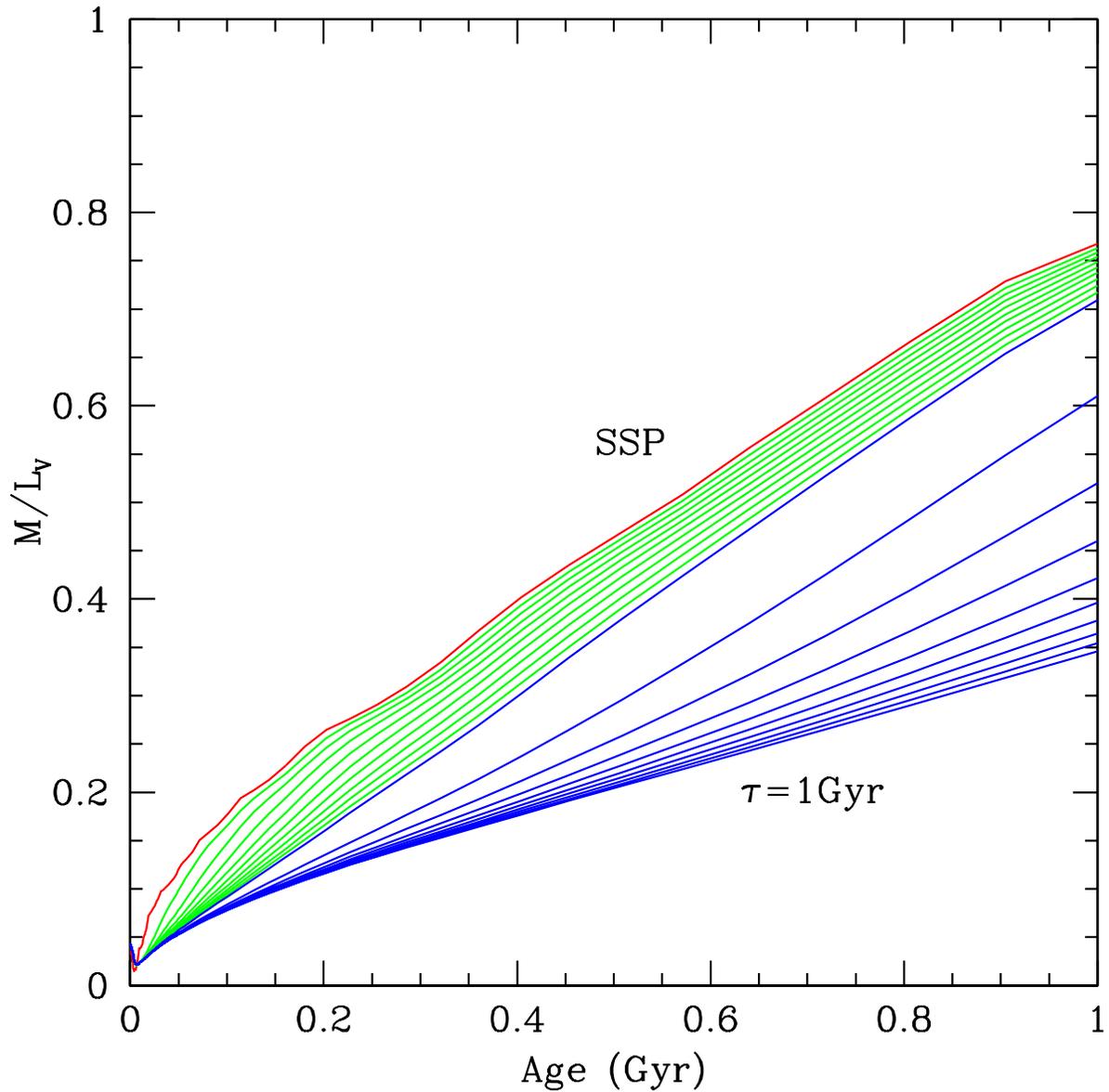}
\caption{This figure demonstrates the evolution of mass-to-light-ratio ($M/L$;
shown here in terms of $M/L_{\rm V}$) for the models considered in our study
(see \S 4). The red curve at the top corresponds to a SSP, the green curves
correspond to models with $\tau$ ranging from 10 to 90~Myr, while the blue
curves correspond to models with $\tau$ from 0.1 to 1.0~Gyr. At the age of 
$\sim$ 1~Gyr, which is the maximum age allowed by our adopted cosmology, a SSP
reaches the largest possible $M/L$ among all the models, and thus gives the
highest possible stellar mass for a given luminosity.
}
\end{figure}

\begin{figure}
%\plotone{fig_age_zch1.eps}
\plotone{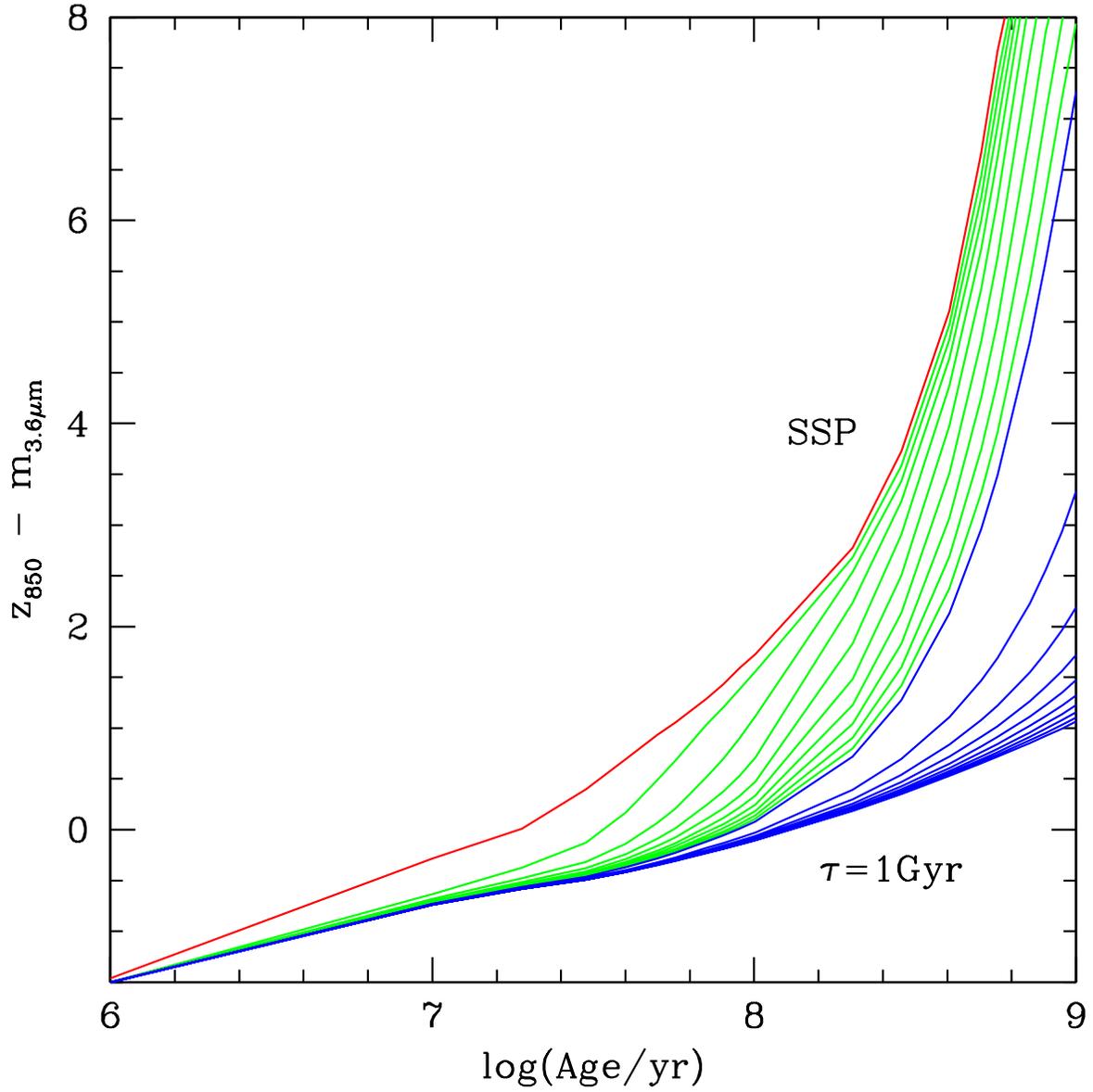}
\caption{The evolution of ($z_{850}-m_{3.6\mu m}$) color for the models
considered in our study (see \S 4). The curves are color-coded in the same
way as in Fig. 2. The ($z_{850}-m_{3.6\mu m}$) color can be used as an age
indicator.
}
\end{figure}

\begin{figure}
%\plotone{fig_mass_hist_all.eps}
\plotone{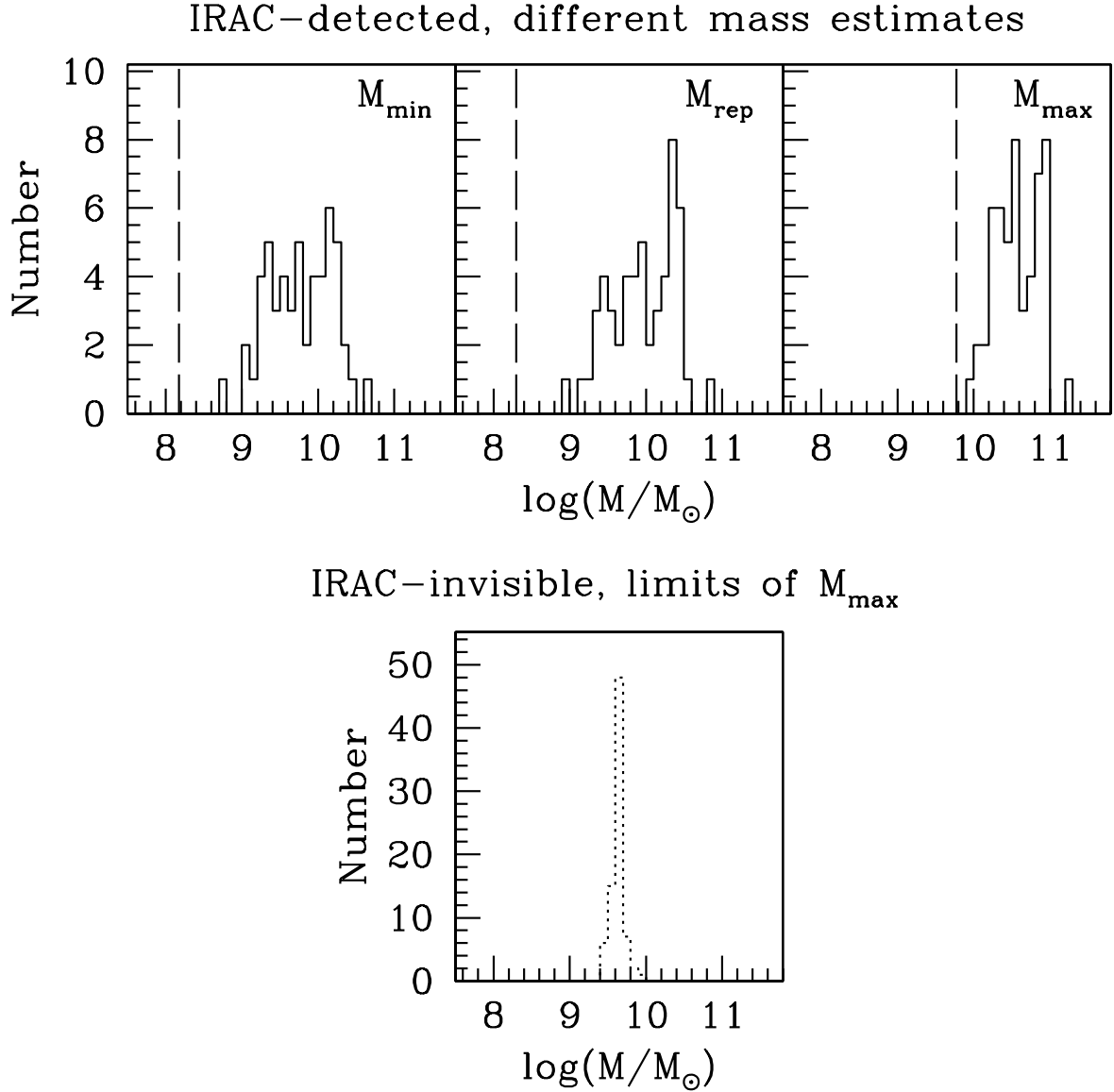}
\caption{The top panel displays the histograms of the stellar mass estimates
of the galaxies in the IRAC-detected samples, using three different
estimators as described in \S 4.1. The vertical dashed line in each figure
represents the corresponding mass estimate of a typical object in the
IRAC-invisible sample (see \S 4.2). The bottom panel shows the histogram of the
upper limits of the stellar masses of the galaxies in the IRAC-invisible
sample, calculated using their flux density upper limits (2~$\sigma$) in the
3.6$\mu$m-channel (see \S 4.2).
}
\end{figure}

\clearpage
\begin{figure}
%\plotone{fig_zch1_z.eps}
\plotone{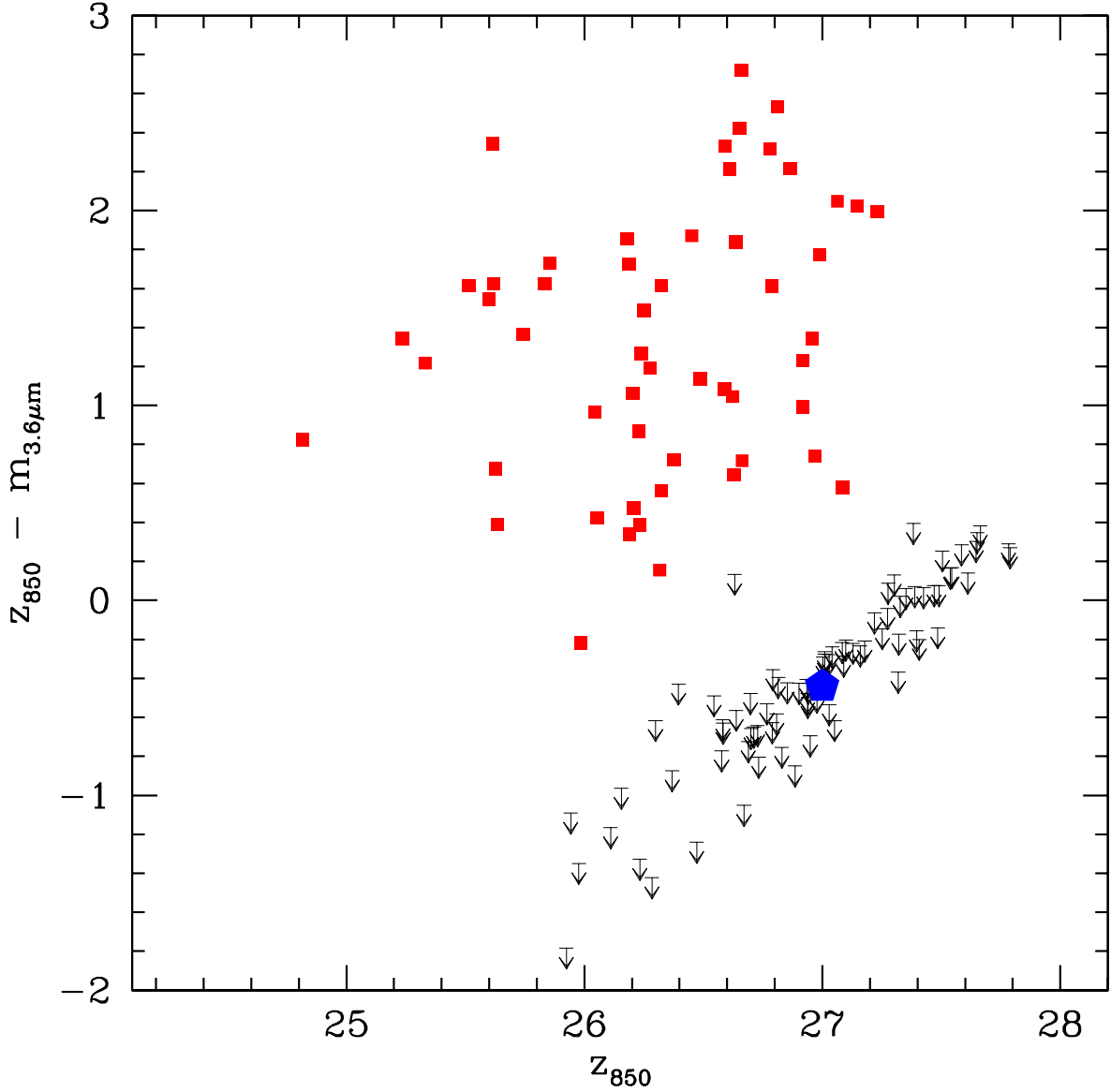}
\caption{The ($z_{850}-m_{3.6\mu m}$) colors of the IRAC-detected objects are
shown as red solid squares, while the upper limits of these colors of the
IRAC-invisible objects are shown as downward arrows. The blue hexagon 
represents the median color of the IRAC-invisible objects, which is calculated
by using the median $z_{850}$ magnitude of the sample and the $m_{3.6\mu m}$
magnitude of the median stack. The IRAC-invisible objects are significantly
bluer than the IRAC-detected ones, indicating that the young populations
are playing a more important role in them than in the IRAC-detected objects.
}
\end{figure}

\clearpage
\begin{figure}
%\plotone{fig_median_stack.eps}
\plotone{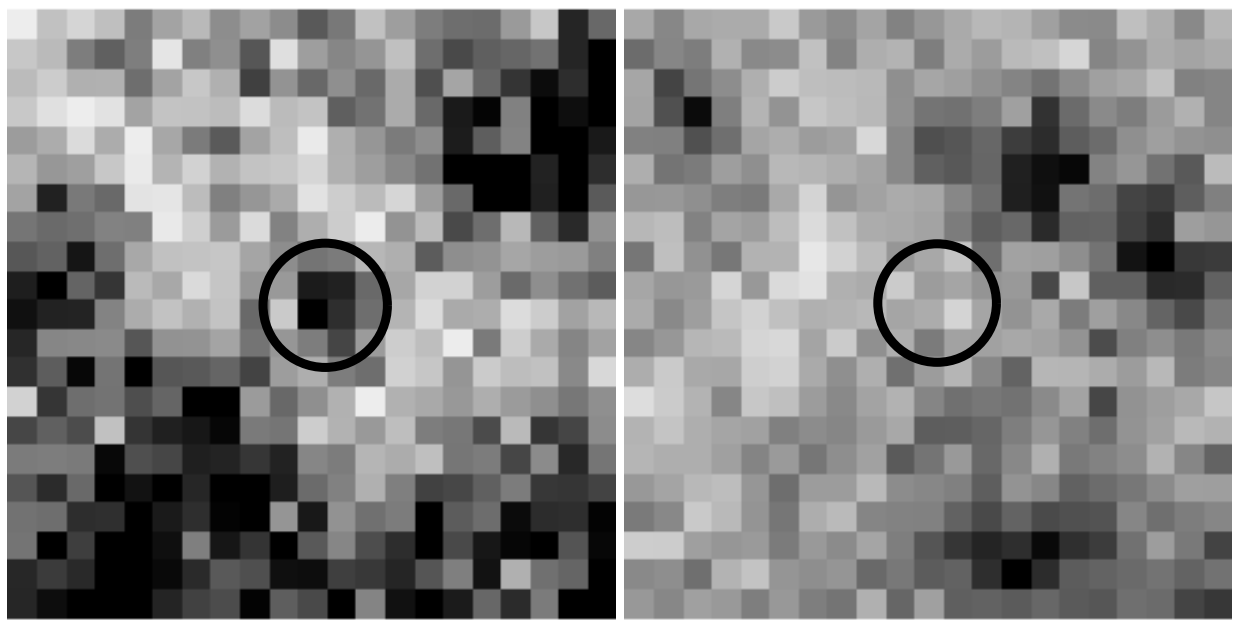}
\caption{The 3.6$\mu m$ median stack of the objects in the
IRAC-invisible sample is displayed in the left panel, where there is a
visible source in the middle (indicated by the circle). For comparison, the
median stacks of the same number of random positions is shown in the right
panel and shows no detected source in the middle circle.
}
\end{figure}

\clearpage
\begin{figure}
%\plotone{fig_compare_z.eps}
\plotone{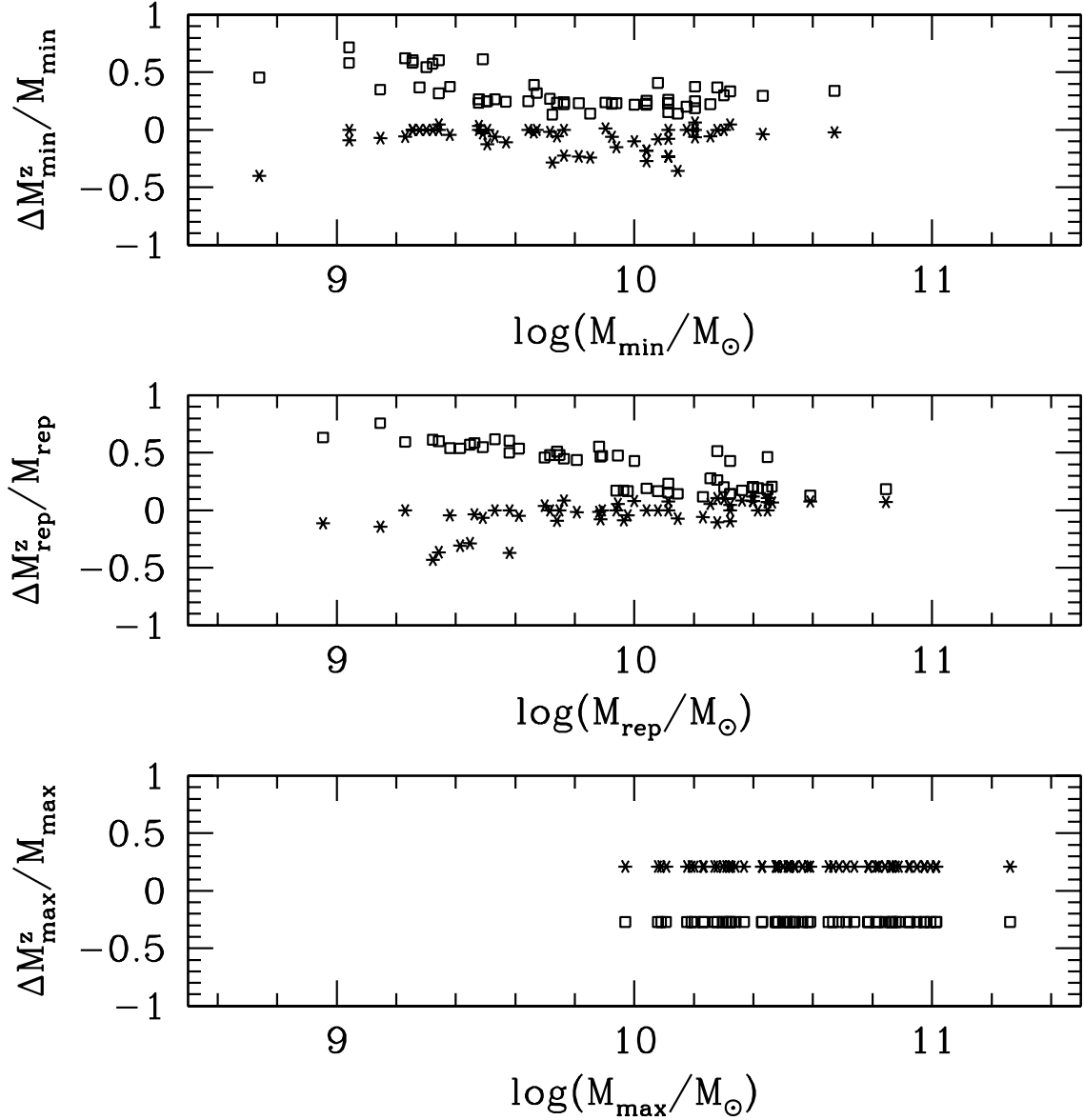}
\caption{The systematic uncertainties of the stellar mass estimates caused
by fixing the redshifts of our models at $z=6$. The evaluations of the three
mass estimates are repeated for every IRAC-detected object by using models at
$z=5.5$ (asterisks) and $z=6.5$ (squares). The systematics are expressed in
terms of relative differences of $M_{\rm min}$, $M_{\rm rep}$ and $M_{\rm max}$
in the top, middle and bottom panels, respectively. We define 
$\Delta M^z=M-M^{'}$, where $M$ is the value obtained with the original models
at $z=6$, and $M^{'}$ is the value obtained with the new models. Thus a 
positive data point means that our original evaluation overestimates the 
quantity if the object is actually at a redshift different from $z=6$, and a
negative data point means the opposite. If our objects are actually all at 
$z=5.5$, our original results underestimate $M_{\rm min}$ and $M_{\rm rep}$
only by a few percent on average, but overestimate $M_{\rm max}$ by 
21\%. If our objects are actually all at $z=6.5$, our original results
overestimate both $M_{\rm min}$ and $M_{\rm rep}$ by $\sim$ 25\%, and 
underestimate $M_{\rm max}$ by 27\%.
}
\end{figure}

\clearpage
\begin{figure}
%\plotone{fig_compare_metal.eps}
\plotone{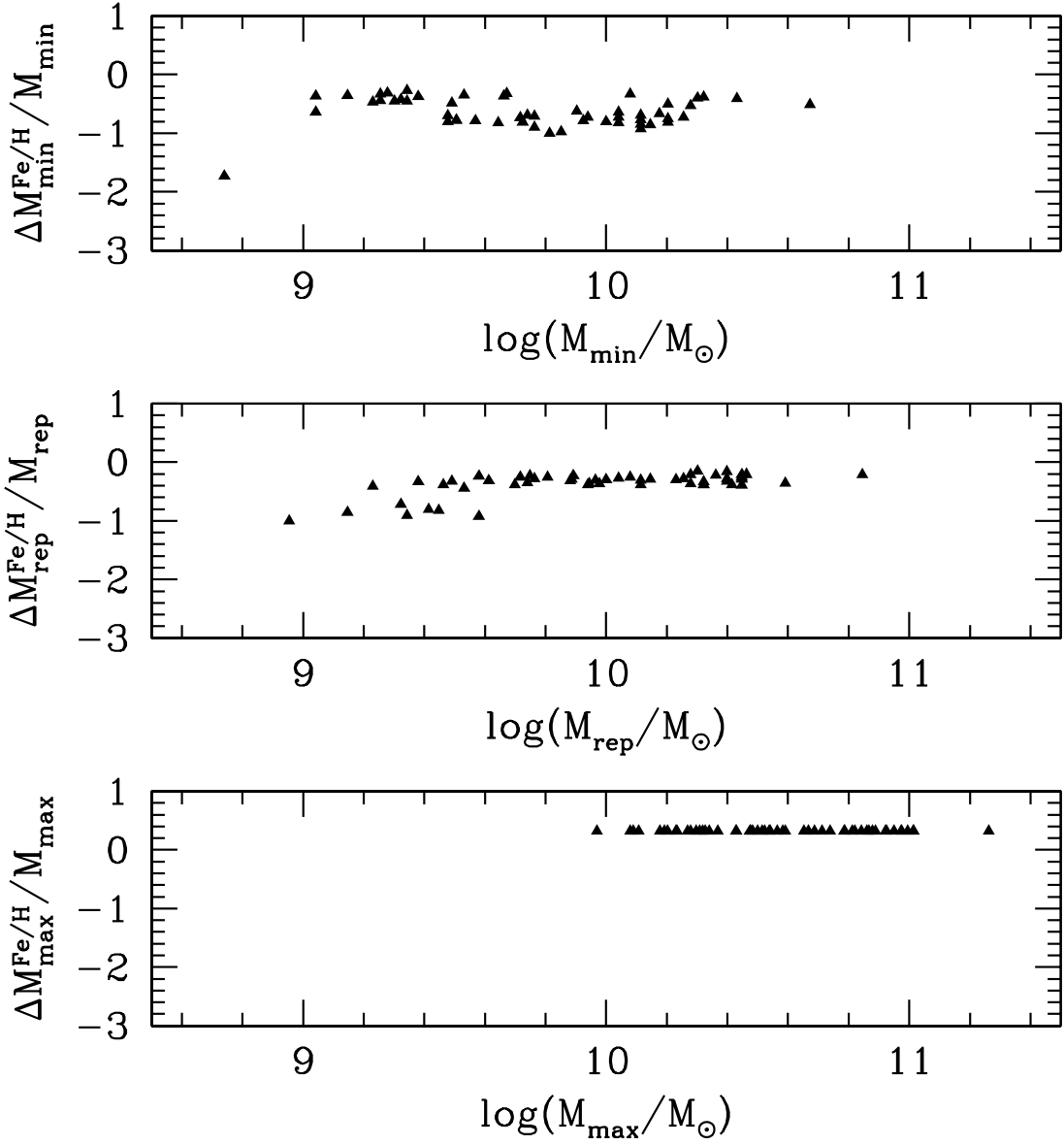}
\caption{Similar to Fig. 7, but for the systematic uncertainties caused by
using models of solar metallicity only. The mass estimates
are recalculated by using models with metallicity of 1/200 of solar, which
is the lowest abundance available for the BC03 models. The relative difference
of mass is defined in a similar way to the quantity shown in Fig. 8. If all
our objects actually have such a low metallicity, our original results
underestimate $M_{\rm min}$ and $M_{\rm rep}$ by $\sim$ 70\% and 32\% on 
average, respectively, and overestimate $M_{\rm max}$ by 32\%.
}
\end{figure}

\clearpage
\begin{figure}
%\plotone{fig_compare_red.eps}
\plotone{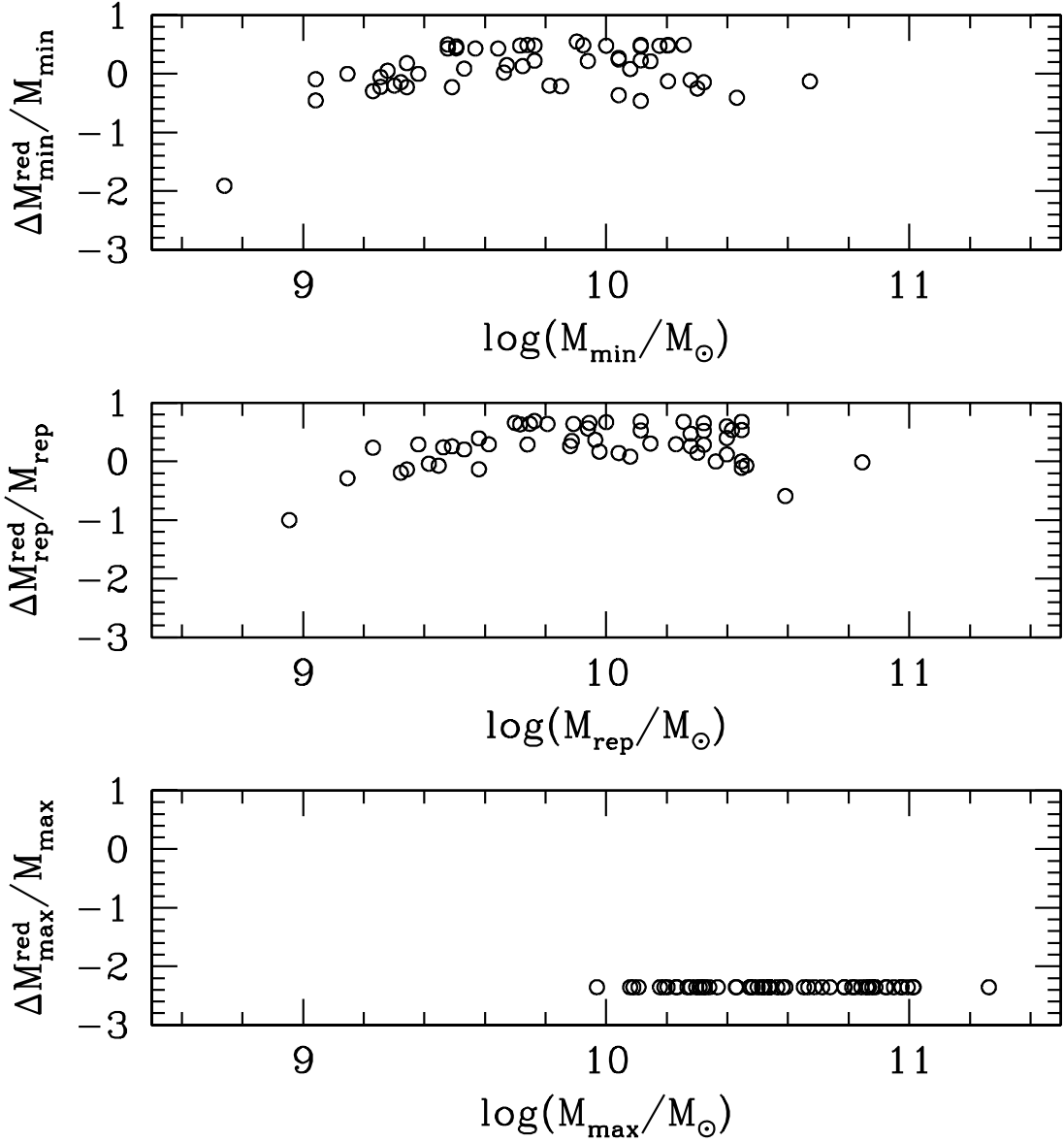}
\caption{Similar to Fig. 7 and 8, but for the systematic uncertainties caused
by using zero-reddening models. The mass estimates are recalculated by using
models with $E(B-V)=0.2$ mag, following the reddening law of Calzetti et al.
(2000). The relative difference of mass is defined in a similar way as before.
If all our objects actually suffer from dust reddening and obscuration of this
amount, our original results overestimate $M_{\rm min}$ and $M_{\rm rep}$ by
$\sim$ 13\% and 30\%, respectively, and underestimate $M_{\rm max}$ by $\sim$ 
a factor of 3.35.
}
\end{figure}

\clearpage
\begin{figure}
%\plotone{fig_specSFR.eps}
\plotone{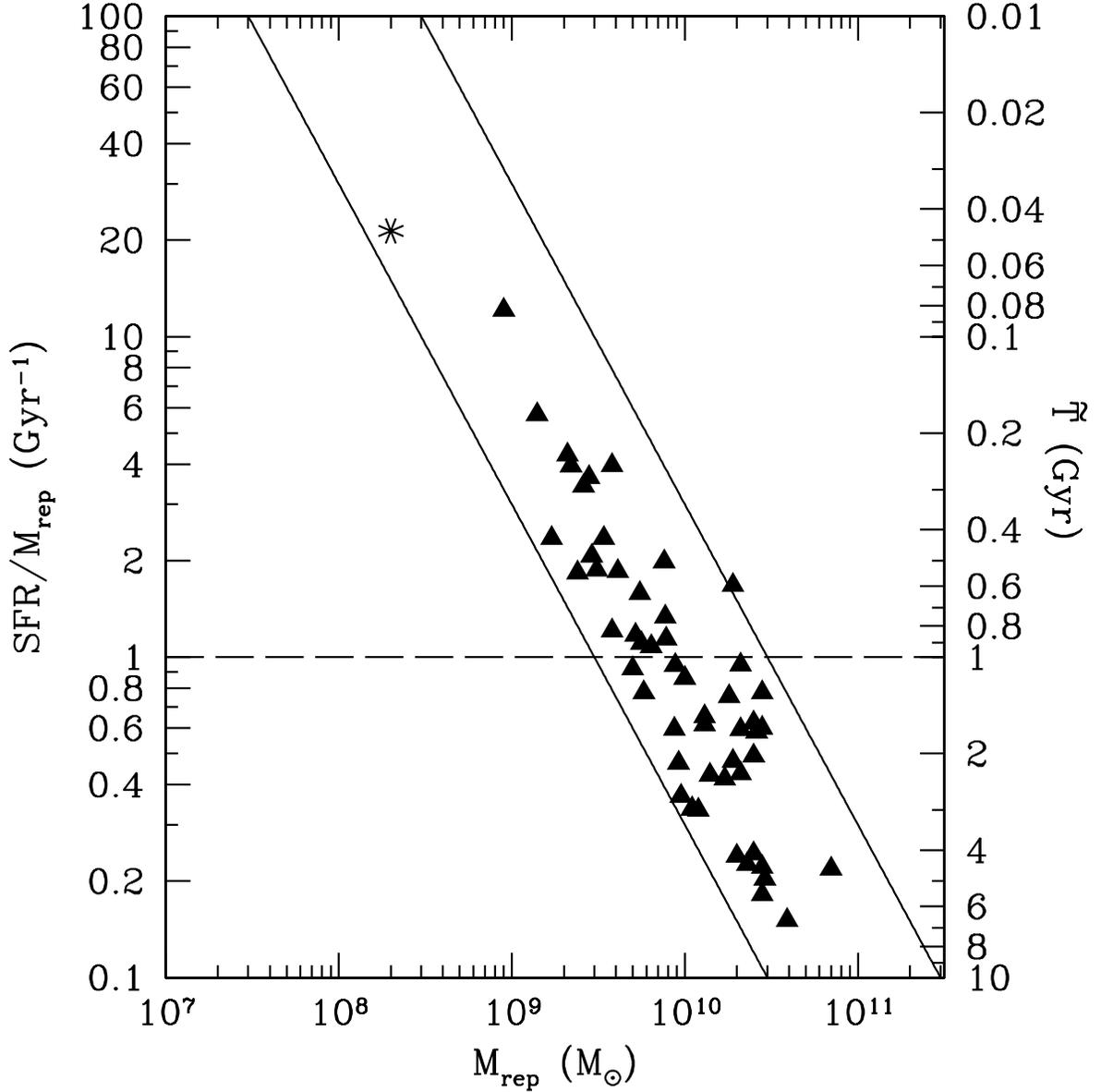}
\caption{The specific SFRs of the IRAC-detected objects (filled triangles) as a
function of $M_{\rm rep}$. For completeness, the specific SFR of the 
IRAC-invisible objects, calculated using their median $z_{850}$ magnitudes and
the median 3.6$\mu m$ stack, is also shown (the asterisk). The two solid lines
shows the specific SFRs of objects with constant SFRs of 3.0 (left) and 
30~$M_\odot$~yr$^{-1}$ (right). The label to the right shows the time periods
through which the galaxies can assemble their total masses if their past SFRs
are the same as the current values. The horizontal dashed line indicates the
age of the universe at $z\approx 6$. Objects below this line must have acquired
their masses at a much higher SFR than the current values.
}
\end{figure}

\clearpage
\begin{figure}
%\plotone{fig_compare2LCDM_cumu.eps}
\plotone{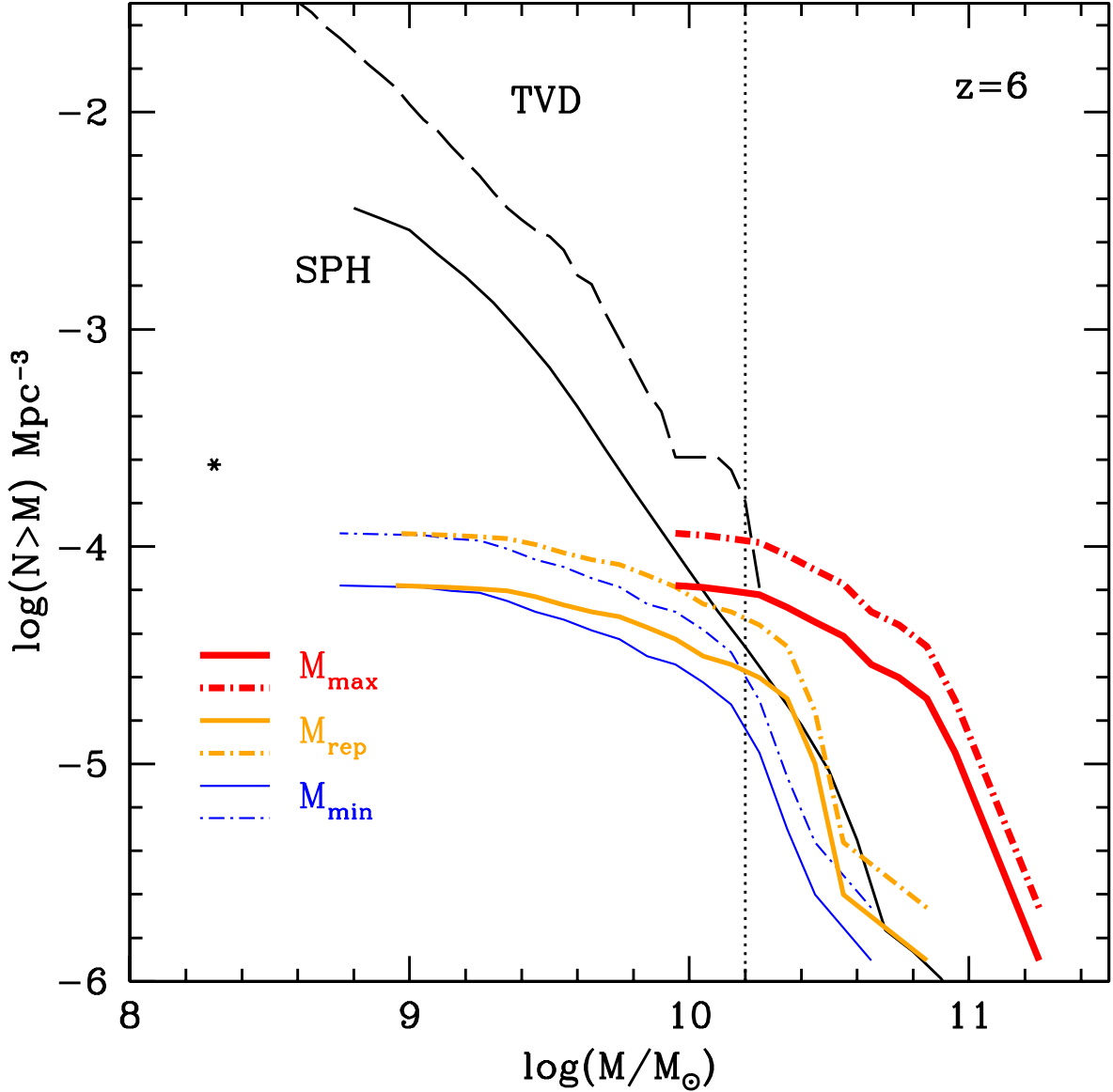}
\caption{The comparison between the observed number density inferred from our
sample and the mass functions predicted by a set of N-body hydrodynamic 
simulations in a $\Lambda$CDM universe from Nagamine et al. (2004) and Night
et al. (2006). Both are presented as cumulative number densities. The 
simulations are shown as the black solid (SPH) and dashed (TVD) curves, while
the observed number densities based on ($M_{\rm min}, M_{\rm rep}, M_{\rm max}$)
are shown as the blue, yellow and red curves, respectively. The solid, 
color-coded curves are the observed values based only on the non-blended,
IRAC-detected sample, while the dot-dashed, color-coded curves are those after
taking the blended sample into account. The asterisk to the left is the
observed cumulative density when assuming all the IRAC-invisible objects
(after correction for blending) have the $M_{\rm rep}$ value as derived using
the median stack (see \S 4.2). The dotted vertical line at 
($M=1.6\times 10^{10}M_\odot$) indicates the mass threshold above which our
IRAC data (but not necessarily the $i_{775}$-dropout sample itself) are
complete.
}
\end{figure}

\clearpage
\begin{figure}
%\plotone{fig_omega_v2.eps}
\plotone{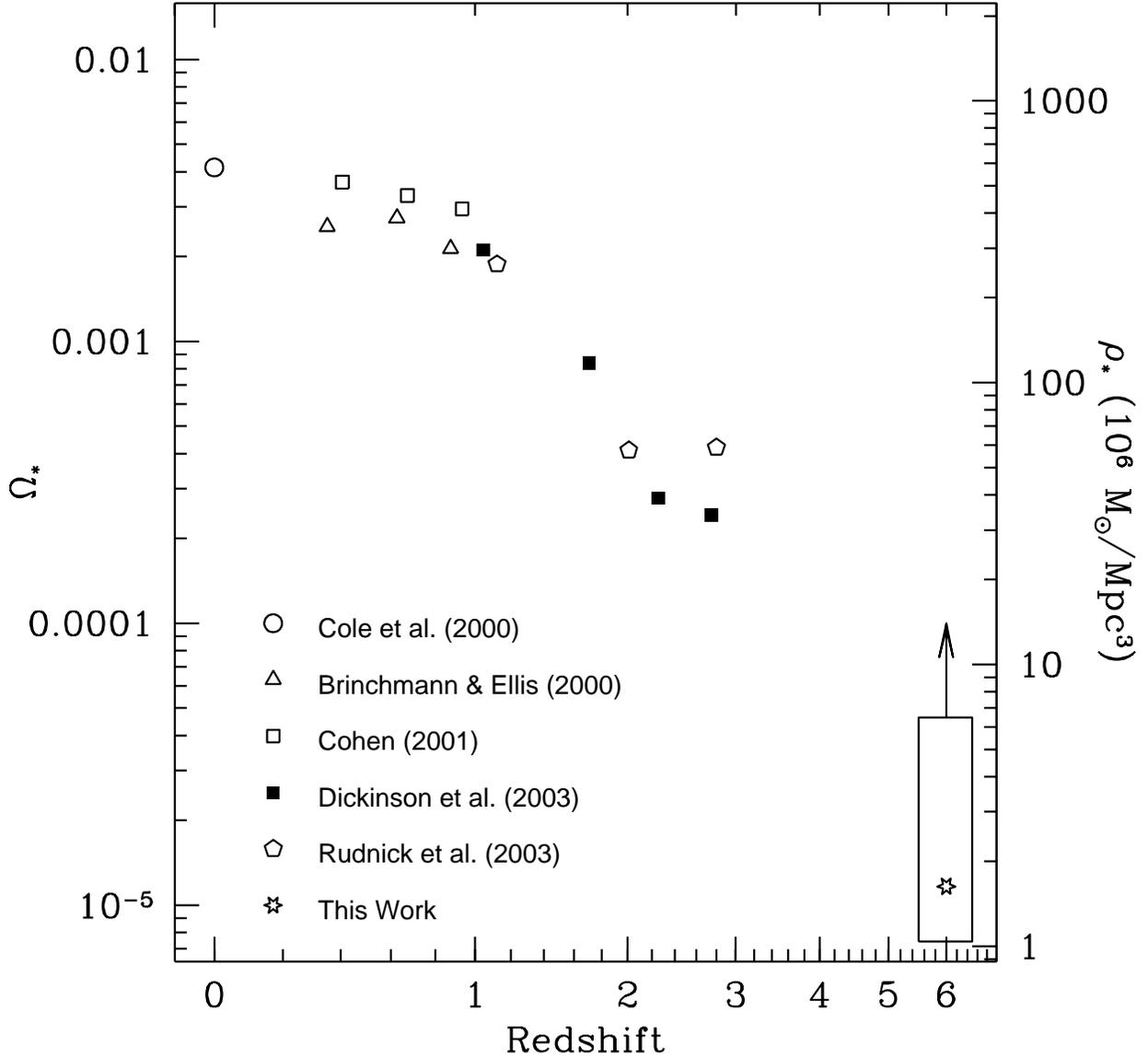}
\caption{The lower limit to the global stellar mass density at $z\approx 6$
based on our study is shown as the rectangle in this figure. The open star
inside this rectangle shows the result based on the $M_{\rm rep}$ values, while
the lower and upper boundaries are set based on the $M_{\rm min}$ and 
$M_{\rm max}$ values, respectively. The upward arrow indicates that these 
results are lower limits. The results at lower redshifts are from Cole et al.
(2000), Brinchmann \& Ellis (2000), Cohen (2001), Dickinson et al. (2003) and
Rudnick et al. (2003). The Y-axis labels to right are in unit of 
$M_\odot/Mpc^3$, while the ones to the left are in terms of ratio to the
critical mass density, which is $1.4\times 10^{11}M_\odot$Mpc$^{-3}$ for our
adopted cosmology.
}
\end{figure}

\clearpage
\begin{figure}
%\plotone{fig_progenitor.eps}
\epsscale{.95}
\plotone{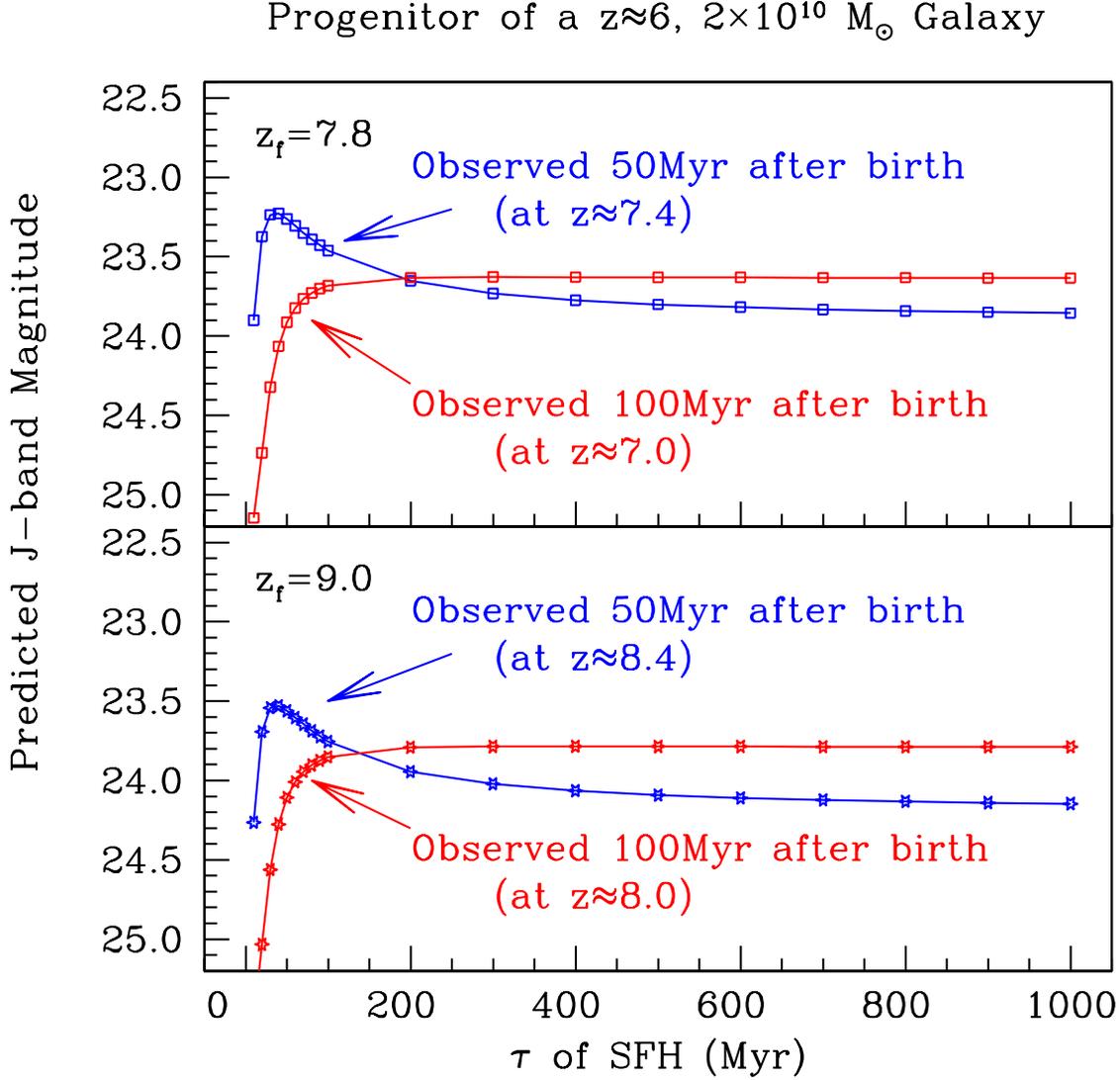}
\epsscale{1}
\caption{The predicted $J$-band brightness of the progenitor of a typical
high-mass object ($M=2\times 10^{10} M_\odot$) in the IRAC-detected sample.
The brightness depends on how quickly the progenitor has been forming stars
(characterized by the timescale $\tau$ of the exponential SFH) and when it
is observed after its birth. Two cases, 50 and 100~Myr after the birth, are
shown as the blue and red curves, respectively. With a formation redshift of
$z_f=7.8$ (top panel), these two ages correspond to being observed at 
$z\approx 7.4$ and 7.0, respectively, while with a formation redshift of
$z_f=9.0$ (bottom panel), they correspond to being observed at $z\approx 8.4$
and 8.0, respectively. These two formation redshifts are chosen for
demonstration because a galaxy evolving to $z\approx 6$ would have an age
of 200--400~Myr, which spans the $T_{\rm rep}$ range of the $M>10^{10}M_\odot$
objects in the IRAC-detected sample. The calculation is done for $\tau=10$~Myr
to 1~Gyr based on the BC03 models, assuming no dust extinction. An modest
amount of dust corresponding to $E(B-V)\approx 0.2$ mag would lower the 
brightness by $\sim$ 2 mag.
}
\end{figure}

\clearpage
\begin{figure}
%\plotone{fig_reion.eps}
\plotone{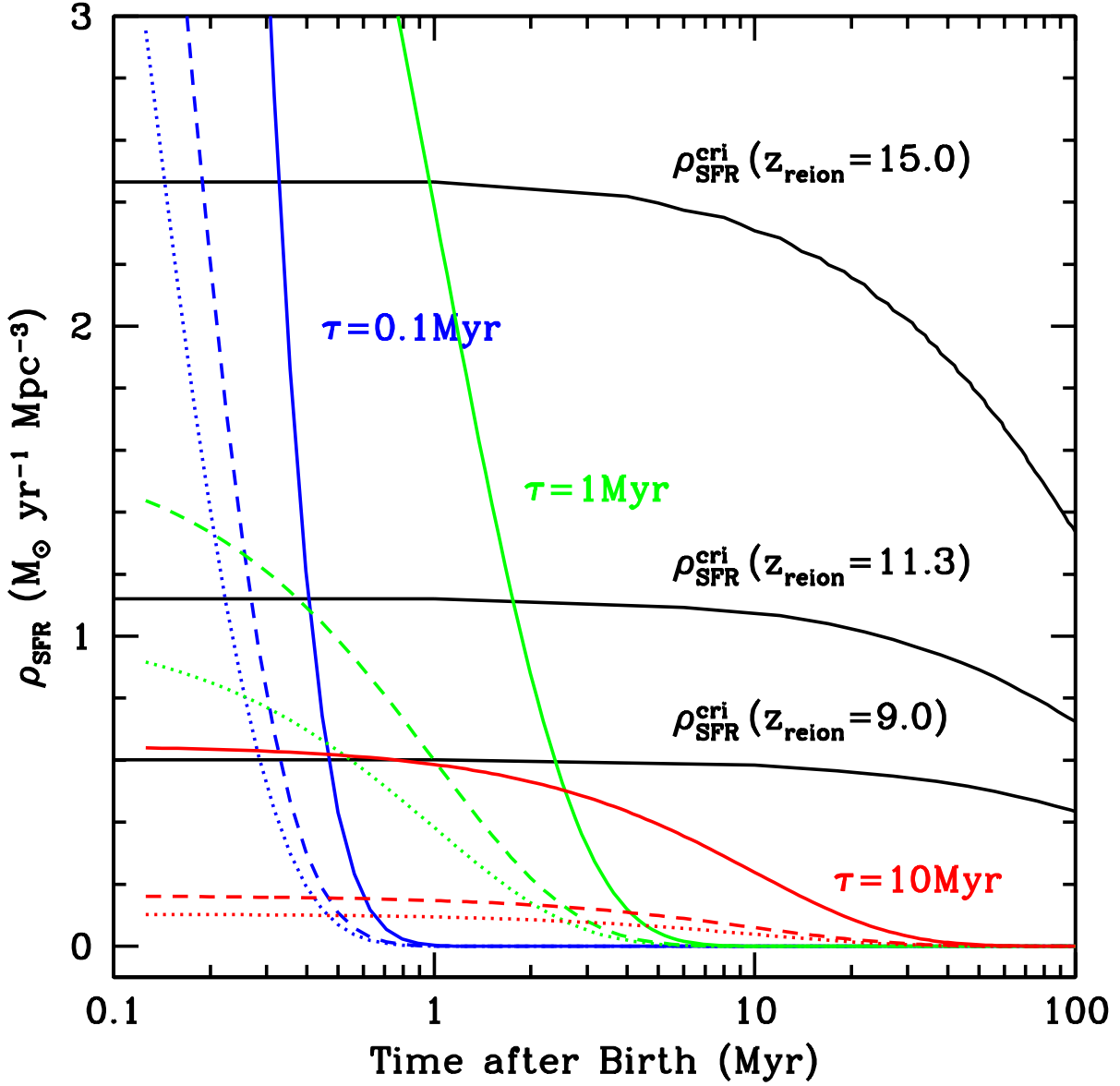}
\caption{The contribution to the reionizing photon background by primeval
galaxies, inferred from the progenitors of the objects in our sample. We
assume an extreme case where the universe started to give birth to all the 
progenitors of the galaxies in our sample at the same time at $z_{\rm reion}$,
and show the contribution of these progenitors to the reionizing photon
background,
expressed in terms of SFR density, as a function of time after the birth. This
contribution is very sensitive to the SFH of the galaxies. Here we show the
results for three exponentially declining SFHs with $\tau$ of 0.1, 1.0 and
10~Myr as blue, green and red curves, respectively. The solid, dashed and
dotted curves represent the results that correspond to the minimum, 
representative and maximum global stellar mass densities shown in Fig. 12,
respectively. The black curves, corresponding to $z_{reion}=15.0$, 11.3 and
9.0, are the critical SFR densities needed to fully ionize the neutral hydrogen
in the universe, calculated according to Madau, Haardt \& Rees (1999). 
See \S 5.2 for detailed interpretation.
}
\end{figure}

\end{document}